\documentclass[aps,prb,groupedaddress,showpacs,twocolumn,superscriptaddress]{revtex4}
\usepackage{graphicx}
\usepackage{amsmath}
\usepackage{amssymb}
\usepackage{bbm}
\usepackage{xspace}
\usepackage{color}
\usepackage{footnote}
\usepackage{comment}
\usepackage{hyperref}

\newcommand{\fig}[1]{Fig.\thinspace{}\ref{#1}}

\newcommand{\Fig}[1]{Figure \ref{#1}}
\newcommand{\eq}[1]{Eq.\thinspace{}(\ref{#1})}

\newcommand{\eqs}[1]{Eqs.\thinspace{}(\ref{#1})}

\newcommand{\se}{Sec.\@\xspace}

\newcommand{\app}{App.\@\xspace}

\newcommand{\tcite}[1]{Ref.~\onlinecite{#1}}
\newcommand{\tcites}[1]{Refs.~\onlinecite{#1}}

\newcommand{\und}{\underline}
\newcommand{\iim}{\Im{}\,}
\newcommand{\li}{\mathcal L}
\newcommand{\ga}[1]{\Gamma^{(#1)}}
\newcommand{\up}{\ensuremath{\uparrow}}
\newcommand{\dw}{\ensuremath{\downarrow}}
\newcommand{\vv}[1]{\boldsymbol{#1}}
\newcommand{\Dph}{\underline{\Delta}_\mathrm{ph}(\omega)}
\newcommand{\Daux}{\underline{\Delta}_\mathrm{aux}(\omega)}

\newcommand{\DRaux}{\Delta^R_\mathrm{aux}(\omega)}

\newcommand{\DKaux}{\Delta^K_\mathrm{aux}(\omega)}

\def\bra#1{\mathinner{\langle{#1}|}}
\def\ket#1{\mathinner{|{#1}\rangle}}
\def\braket#1{\mathinner{\langle{#1}\rangle}}

\newcommand{\nag}{{\phantom{\dagger}}}

\newcommand{\pim}{IM$_\mathrm{ph}$ }
\newcommand{\aim}{IM$_\mathrm{aux}$ }
\newcommand{\pimns}{IM$_\mathrm{ph}$}
\newcommand{\aimns}{IM$_\mathrm{aux}$}

\begin{document}

\title{Auxiliary master equation approach within matrix product states:\\Spectral properties of the nonequilibrium Anderson impurity model.}

\author{Antonius Dorda}
\email[]{dorda@tugraz.at}
\affiliation{Institute of Theoretical and Computational Physics, Graz University of Technology, 8010 Graz, Austria}
\author{Martin Ganahl}
\affiliation{Institute of Theoretical and Computational Physics, Graz University of Technology, 8010 Graz, Austria}
\affiliation{Perimeter Institute for Theoretical Physics, Waterloo, Ontario}
\author{Hans Gerd Evertz}
\affiliation{Institute of Theoretical and Computational Physics, Graz University of Technology, 8010 Graz, Austria}
\author{Wolfgang von der Linden}
\affiliation{Institute of Theoretical and Computational Physics, Graz University of Technology, 8010 Graz, Austria}
\author{Enrico Arrigoni}
\email[]{arrigoni@tugraz.at}
\affiliation{Institute of Theoretical and Computational Physics, Graz University of Technology, 8010 Graz, Austria}

\date{\today}

\begin{abstract}
Within the recently introduced auxiliary master equation approach it is possible to address steady state properties of strongly correlated impurity models, small molecules or clusters efficiently and with high accuracy. It is particularly suited for dynamical mean field theory in the nonequilibrium as well as in the equilibrium case. The method is based on the solution of an auxiliary open quantum system, which can be made quickly equivalent to the original impurity problem. In its first implementation a Krylov space method was employed. Here, we aim at extending the capabilities of the approach by adopting matrix product states for the solution of the corresponding auxiliary quantum master equation. This allows for a drastic increase in accuracy and permits us to access the Kondo regime for large values of the interaction. In particular, we investigate the nonequilibrium steady state of a single impurity Anderson model and focus on the spectral properties for temperatures $T$ below the Kondo temperature $T_
K$ and for small bias voltages $\phi$. For the two cases considered, with $T\approx T_K/4$ and $T\approx T_K/10$ we find a clear splitting of the Kondo resonance into a two-peak structure for $\phi$ close above $T_K$.  In the equilibrium case ($\phi=0$) and for $T\approx T_K/4$, the obtained spectral function essentially coincides with the one from numerical renormalization group. 
\end{abstract}
\pacs{71.15.-m,71.27+a,72.15.Qm,73.21.La,73.63.Kv}

\maketitle

\section{Introduction}\label{sec:introduction}
The equilibrium properties of the single impurity Anderson model (SIAM) and the associated Kondo model,~\cite{ande.61,kond.64,sc.wo.66} originally devised in the process of investigating metal hosts with dilute magnetic impurities,~\cite{ha.bo.34,wils.53,sa.co.64} are nowadays well understood.~\cite{hews.93,bu.co.08} Renormalization group (RG) methods provided first perturbative analyses,\cite{ande.70} and especially the development of Wilson's numerical RG (NRG)~\cite{wils.75} allowed to properly capture the universal low-energy physics, governed by an exponentially small energy scale, the Kondo temperature $T_K$.~\cite{bu.co.08} The field of correlated impurity models has gained renewed interest due to novel experimental realizations in quantum dots,~\cite{go.sh.98,wi.fr.00,fr.ha.02,le.sc.05,fr.wi.10} single molecule transistors,~\cite{cu.fa.05,ag.ye.03,sm.no.02,pa.ab.02,li.sh.02} and from a theoretical point of view, due to its importance for dynamical mean field theory 
(DMFT).~\cite{me.vo.89,ge.ko.92, ge.ko.96,ko.sa.06,voll.12,gu.mi.11,bu.co.08} 
The extension of DMFT to the nonequilibrium case can be carried out within the Keldysh formalism.~\cite{fr.tu.06,sc.mo.02u,ao.ts.14} Nonequilibrium  DMFT and different applicable impurity solvers have been thoroughly discussed in other work, see for instance \tcites{ao.ts.14,okam.07,okam.08,jo.fr.08,ec.ko.09,gr.ba.13,ar.kn.13,do.nu.14,ti.do.15u}. 

In the present study we want to focus on the physics of the impurity problem out of equilibrium itself, with an implementation of  the auxiliary master equation approach (AMEA)~\cite{ar.kn.13,do.nu.14} based on matrix product states (MPS). Already in a first study, where Krylov space methods were employed,~\cite{do.nu.14} AMEA has proven to feature a systematically improvable accuracy and to yield a well-defined Kondo peak in equilibrium together with a splitting in the nonequilibrium case. However, the exponential scaling of Krylov space methods with system size sets a ``hard limit'' to the achievable accuracy, and thus to the lowest temperatures accessible. The MPS extension presented here turns out to be crucial in order to achieve highly accurate results in the Kondo regime down to low temperatures and up to large interactions. In the equilibrium case, the accuracy of our results becomes even comparable to NRG.

Specifically, we investigate the nonequilibrium steady state dynamics of a SIAM, which is driven by the coupling to two leads at different chemical potentials, caused by an external bias voltage $\phi$. Impurity models in such a setup were considered already by many groups, numerically as well as analytically.~\cite{an.me.10,hers.93,pa.na.06} To give a brief non-exhaustive overview, different techniques employed are the noncrossing approximation,~\cite{wi.me.94,le.sc.01,ro.kr.01} real-time diagrammatic methods,~\cite{ko.sc.96} Keldysh perturbation theory,~\cite{fu.ue.03} Keldysh effective field theory,~\cite{sm.gr.11,sm.gr.13} dual fermions,~\cite{ju.li.12,mu.bo.13} perturbative RG,~\cite{ro.pa.03,sh.ro.06} flow equations,~\cite{kehr.05,fr.ke.10} functional RG,~\cite{ge.pr.07,ja.pl.10b} real-time RG,~\cite{sa.we.12,pl.sc.12,re.pl.14,an.sc.14} time-dependent density matrix RG,~\cite{he.fe.09,ho.mc.09,nu.ga.13,nu.ga.15} NRG,~\cite{ande.08,sc.an.11,jo.an.13} Monte Carlo methods,~\cite{we.ok.10,ha.he.07,co.gu.14}
 as well as cluster approaches.~\cite{nu.he.12} The properties of the correlated impurity have been established in certain limits, for example for high temperatures $T\gg T_K$ or high biases $\phi\gg T_K$, where the Kondo effect is strongly suppressed by decoherence and the problem reduces to a weak coupling one.~\cite{ro.kr.01,pa.ro.04,fr.ke.10,pl.sc.12,ka.na.99,co.ho.01} A splitting of the Kondo peak in the spectral function was found at sufficiently high bias voltages and low $T$, with a two-peak structure pinned to the chemical potentials of the leads.~\cite{wi.me.94,le.sc.01,ro.kr.01,ko.sc.96,fu.ue.03,sh.ro.06,fr.ke.10,nu.he.12,ande.08,ha.he.07,co.gu.14} In the other limit $\phi\ll T_K$ and $T\ll T_K$, linear response as well as Fermi liquid theory is applicable.~\cite{mu.bo.13,re.pl.14,ogur.05,se.ma.09} Nevertheless, the intermediate and low-energy nonequilibrium regime, where both $T$ and $\phi$ are of the order of and especially below $T_K$, remains challenging and the spectral properties could not 
yet be completely resolved. 
Work in this direction has for example been done in \tcites{ande.08,ha.he.07,co.gu.14}. However, the extension of NRG to the nonequilibrium case still leaves open questions,~\cite{rosc.11} and the Monte Carlo approaches, even though numerically exact, are either limited to relatively high temperatures and short times, or involve a demanding double analytical continuation.~\cite{gu.re.10,di.we.10} With the work presented here, we want to contribute to these findings and present well-resolved spectral data for cases where both $T\lesssim T_K$ and $\phi\lesssim T_K$.

\section{Model and method}\label{sec:amea}
The basic idea of AMEA is to map a general correlated impurity model in or out of equilibrium, here referred to as the physical impurity model (\pimns), onto an appropriately chosen auxiliary one (\aimns), which is small enough to be solvable precisely by numerical techniques. The self-energy of \aim serves then as an approximation to the one of \pimns. Specifically, \aim is modeled by an open quantum system described by a Lindblad equation, which consists of a finite number of bath sites and additional Markovian environments. In the mapping procedure, the bath parameters of \aim are optimized in order to reproduce the dynamics of \pim as closely as possible. By increasing the number of bath sites $N_B$, more optimization parameters are available and a convergence (typically exponential) towards the exact solution of \pim is achieved. The mapping is formulated in terms of the hybridization function of \aimns, which is obtained through a single-particle calculation, and the many-body problem is solved 
thereafter.

AMEA itself and a solution strategy for the correlated \aim based on exact diagonalization (ED) was presented in detail in \tcites{ar.kn.13,do.nu.14}. Here, we make use of MPS in order to solve for the correlation functions, which enables us to treat auxiliary systems with a larger number of bath sites. In the following we briefly summarize the governing equations in AMEA and point out modifications in the construction of \aim favorable for an MPS treatment. After that, the MPS implementation is discussed.

\subsection{Keldysh Green's functions}\label{ssec:negf}
In general, for nonequilibrium situations Green's functions are conveniently defined on the Keldysh contour.~\cite{schw.61,kad.baym,keld.65,ra.sm.86,ha.ja,wagn.91} Since we are particularly interested in the long-time limit, where a steady state is reached, time translational invariance applies and the Keldysh Green's functions can be written in the frequency domain
\begin{align}
  \und{G}(\omega) &= \begin{pmatrix} G^R(\omega) & G^K(\omega) \\ 0 & G^A(\omega) \end{pmatrix}\,,
  \label{eq:negf}
\end{align}
with $G^A = (G^R)^\dagger$, and we denote by an underscore $\und{\cdots}$ a $2\times2$ object in Keldysh space. Only in an equilibrium situation the Keldysh component is related to the retarded one via the fluctuation dissipation theorem
\begin{equation}
 G^K(\omega) = 2i\left(1-2 p_\mathrm{FD}(\omega,\mu,T)\right) \iim\left\{G^R(\omega) \right\}\,,
 \label{eq:FDT}
\end{equation}
where $p_\mathrm{FD}(\omega,\mu,T)$ represents the Fermi-Dirac distribution.
In contrast, in a general nonequilibrium situation a distribution function is not known a priori and the Keldysh and the retarded component have to be considered as independent functions.

It is convenient to introduce the steady state lesser and greater Green's functions
\begin{equation}
 G^<(t) = i\braket{c^\dagger(t)c} \,,\hspace{1em} G^>(t) = -i\braket{c(t)c^\dagger} \,,
 \label{eq:Gless}
\end{equation}
for generic fermionic creation/annihilation operators $c^\dagger/c$, which are related to $G^R$ and $G^K$ by
\begin{align}
 G^R(\omega)-G^A(\omega) &= G^>(\omega) - G^<(\omega)  = -2i\pi A(\omega) \,,  \nonumber \\
 G^K(\omega) &= G^>(\omega) + G^<(\omega) \,,
\end{align}
and $A(\omega)$ is the spectral function. Throughout this work we consider solely steady state expectation values and denote them in compact notation by $\braket{\dots}$, cf. \eq{eq:Gless}.

\subsection{Physical impurity model}\label{ssec:im_ph}
In this work, we consider for \pim a single impurity Anderson model in a nonequilibrium setup, given by an impurity Hamiltonian $ H_{\mathrm{imp}}$, two noninteracting fermionic leads representing the electronic reservoir $ H_{\mathrm{res}}$, and an impurity-reservoir coupling $ H_{\mathrm{coup}}$:
\begin{equation}
  H_{\mathrm{ph}} =  H_{\mathrm{imp}} +  H_{\mathrm{res}} +  H_{\mathrm{coup}}\,.
 \label{eq:H_ph}
\end{equation}
The correlated impurity consists of a single level with energy $\varepsilon_d$ and on-site Hubbard interaction $U$  
\begin{equation}
 H_{\mathrm{imp}} = \varepsilon_d \sum_{\sigma\in\{\uparrow,\downarrow\}} d^\dagger_{\sigma} d_{\sigma}^\nag + U \left(d^\dagger_{\up}d^\nag_{\up}-\frac{1}{2}\right) \left(d^\dagger_{\dw}d^\nag_{\dw}-\frac{1}{2}\right)\,,
 \label{eq:H_imp}
\end{equation}
where $d^\dagger_\sigma$/$d^\nag_\sigma$ are fermionic creation and annihilation operators on the impurity site. The reservoir Hamiltonian can be written in terms of the energy levels $\varepsilon_{\lambda k}$ and potentials $\varepsilon_\lambda$ for the two leads $\lambda$
\begin{equation}
 H_{\mathrm{res}} = \sum_{\lambda \in \{L,R\}} \sum_{k \sigma} \left(\varepsilon_\lambda + \varepsilon_{\lambda k} \right) a^\dagger_{\lambda k \sigma}a^\nag_{\lambda k \sigma} \,,
 \label{eq:H_res}
\end{equation}
and the impurity-reservoir coupling is given by
\begin{equation}
 H_{\mathrm{coup}} = \frac{1}{\sqrt{N_k}}\sum_{\lambda k \sigma} t'_\lambda \left( a^\dagger_{\lambda k \sigma}d^\nag_{\sigma} + \mathrm{h.c.} \right) \,,
 \label{eq:H_coup}
\end{equation}
with $a^\dagger_{\lambda k \sigma}$/$a^\nag_{\lambda k \sigma}$ representing creation and annihilation operators for lead electrons.

Throughout this work we consider the particle-hole symmetric case with $\varepsilon_d=0$, $t'_L= t'_R$, $\varepsilon_{L k} = \varepsilon_{R k}$. An externally applied bias voltage $\phi$ results in an anti-symmetrical shift of the chemical potentials $\mu_{L/R} = \pm\frac{\phi}{2}$. In \se\ref{ssec:results_Lorentzian} we further consider for the on-site energies the case $\varepsilon_{L/R} = \pm\frac{\phi}{2}$, whereas in the rest of the work the voltage does not shift the lead energies. This is irrelevant for $\phi$ much smaller than the bandwidth.

The Green's function of \pim is given by the Dyson equation
\begin{equation}
\underline{G}_\mathrm{ph}^{-1}(\omega)=\underline{g}^{-1}_{0}(\omega) - \underline{\Delta}_\mathrm{ph}(\omega)-\underline{\Sigma}_\mathrm{ph}(\omega)\,.
\label{eq:dyson_ph}
\end{equation}
Here, $\underline{g}_{0}$ denotes the noninteracting Keldysh Green's function of the decoupled impurity, i.e. $g^R_0 = \left(\omega - \varepsilon_d \right)^{-1}$ and $(\underline{g}^{-1}_{0})^K$ can be neglected. The hybridization function $\underline{\Delta}_\mathrm{ph}$ is given by the sum of contributions from the two leads
\begin{equation}
\underline{\Delta}_\mathrm{ph}(\omega) = \sum_\lambda {t'_\lambda}^2 \underline{g}_\lambda(\omega)\,,
\label{eq:Delta}
\end{equation}
where $\underline{g}_\lambda(\omega)$ denote lead Green's functions at the contact point in the decoupled case. Except in the calculations presented in \se\ref{ssec:results_Lorentzian}, we consider throughout this work a flat band model with the retarded component of $\underline{g}_\lambda(\omega)$ given by
\begin{equation}
 -\iim\{g^R_\lambda(\omega)\} =  \frac{\pi}{2D} \Theta\left(D-|\omega| \right)\,,
 \label{eq:gR_lambda}
\end{equation}
where we choose the hybridization strength $\Gamma =  {t'_\lambda}^2 \pi/D$ as unit of energy and take $D=10\,\Gamma$. The real part is determined via the Kramers Kronig relation. For the fit in the mapping procedure (see \se\ref{ssec:mapping}) it is of advantage to deal with smooth functions of $\omega$, so that we introduce in \eq{eq:gR_lambda} a smearing of the cut-offs in the Heaviside function, determined by Fermi functions $p_\mathrm{FD}(\omega,\pm D,0.5\,\Gamma)$ with an artificial temperature $0.5\,\Gamma$. Since this modification is well outside the scale of the impurity energies, it does not affect the low energy physics.

The decoupled leads are in equilibrium, so that the Keldysh component $g^K_\lambda(\omega)$ of each lead is given by \eq{eq:FDT} with the corresponding chemical potential $\mu_\lambda$. The temperature $T$ is taken to be the same in both of the leads.
Notice that the Keldysh component is the only $T$-dependent quantity and results for different $T$ shown below differ only in the smearing of the Fermi edge in $g^K_\lambda(\omega)$. In particular, we are interested in temperatures close to and below the Kondo temperature $T_K$. As for other methods, the low-temperature regime is most challenging (cf. \se\ref{ssec:mapping} and \app\ref{app:conv_NB}). For a  Hubbard interaction of $U = 6\,\Gamma$, as considered throughout the work, one finds for the flat band model $T_K \approx 0.2\,\Gamma$.~\cite{footnote_nrgTk,footnote_nrg,bu.co.08,os.zi.13,ha.we.14}

The remaining unknown quantity in \eq{eq:dyson_ph} is the self-energy $\underline{\Sigma}_\mathrm{ph}(\omega)$, which cannot be determined exactly since \pim is interacting and of infinite size. This is evaluated by means of the mapping to \aimns.

\subsection{Auxiliary impurity model}\label{ssec:im_aux}
For \aim we take an open quantum system of finite size, embedded in Markovian environments and described by a Lindblad equation for the system density operator $\rho$:
\begin{equation}
 \frac{d}{dt}\rho = \li \rho\,.
 \label{eq:Leq}
\end{equation}
The Lindblad super-operator $\li = \li_H + \li_D$ consists of an unitary part $\li_H\rho = -i[H_{\mathrm{aux}},\rho]$ and the dissipator $\li_D$ as described below.~\cite{footnote_comm}

Additionally to the original impurity site we consider $N_B$ bath sites arranged in a linear geometry. For convenience we choose $N_B$ even and the impurity site at the center, specified by the index $f$. The Hamiltonian for \aim is given by
\begin{equation}
   H_{\mathrm{aux}} = \sum_{ij\sigma} E_{ij} c_{i\sigma}^\dagger c^\nag_{j\sigma} + U n_{f\up} n_{f\dw} \,.
 \label{eq:H_aux}
\end{equation}
Here $n_{f\sigma} = c_{f\sigma}^\dagger c^\nag_{f\sigma}$ with $c_{i\sigma}^\dagger$/$c^\nag_{i\sigma}$ the fermionic creation/annihilation operators and the $(N_B+1)\times(N_B+1)$ matrix $\vv{E}$ couples only nearest neighbor (n.n.) terms, i.e. it is tridiagonal in the chosen geometry. To end up with a noninteracting bath we allow at most for Lindblad operators that are linear in $c_{i\sigma}^\dagger$/$c^\nag_{i\sigma}$. The dissipator is then given by~\cite{footnote_comm}
\begin{align}
   \li_D \rho &= 2 \sum_{ij\sigma} \ga1_{ij} \left( c^\nag_{j\sigma}\rho c_{i\sigma}^\dagger - \frac{1}{2} \left\{\rho, c_{i\sigma}^\dagger c^\nag_{j\sigma} \right\} \right)  \nonumber \\
   &\hspace{1em}+ 2 \sum_{ij\sigma} \ga2_{ij} \left( c_{i\sigma}^\dagger\rho c^\nag_{j\sigma}  - \frac{1}{2} \left\{\rho, c^\nag_{j\sigma}c_{i\sigma}^\dagger \right\} \right)   \,.
 \label{eq:L_D}
\end{align}
Both matrices of coupling constants $\vv{\ga1}$ and $\vv{\ga2}$ are symmetric and positive definite.~\cite{footnote_gamma} 

A key aspect in AMEA is that the bath parameters in the Lindblad equation are not determined within conventional Born-Markov approximations,~\cite{br.pe.02,scha.14,nu.do.15u} but are only used as fit parameters to optimally reproduce $\Dph$ by $\Daux$, see \se\ref{ssec:mapping}.

Once the parameters of \aim are determined, the many-body problem is solved (cf. \se\ref{ssec:manybody}) in order to obtain the interacting Green's function
\begin{equation}
\und{G}_\mathrm{aux}^{-1}(\omega)=\und{g}^{-1}_{0}(\omega) - \und{\Delta}_\mathrm{aux}(\omega)-\und{\Sigma}_\mathrm{aux}(\omega)\,.
\label{eq:dyson_aux}
\end{equation}
At this point it is convenient to set $\und{\Sigma}_\mathrm{ph}(\omega) = \und{\Sigma}_\mathrm{aux}(\omega) = \und{\Sigma}(\omega)$, so that we obtain from \eq{eq:dyson_ph} a very accurate result for the Green's function of \pimns. In this way, the $U=0$ limit is recovered exactly.

\subsection{Mapping procedure}\label{ssec:mapping}
In order to have a faithful representation of the dynamics of \pim by \aimns, we need to fulfill $\Daux \approx \Dph$ as closely as possible. For local quantities and correlation functions on the impurity, the influence of the bath is completely determined by the hybridization function only, independently of the specific bath geometry. Therefore, the mapping becomes exact in the limit $\Daux \equiv \Dph$. To achieve $\Daux \approx \Dph$, we minimize the mean squared error between them as a function of the bath parameters in the Lindblad equation, i.e. the matrices $\vv{E}$, $\vv{\ga1}$ and $\vv{\ga2}$.

It is important to stress that a single particle calculation is sufficient to determine $\Daux$, for which the Green's functions read~\cite{ar.kn.13,do.nu.14}
\begin{align}
 \vv{G}_0^R(\omega) &= \left(\omega - \vv{E} +i \left(\vv{\ga1}+\vv{\ga2}\right)  \right)^{-1} \,,\nonumber\\
 \vv{G}_0^K(\omega) &= 2i \vv{G}_0^R(\omega) \left(\vv{\ga2} - \vv{\ga1} \right) \vv{G}_0^A(\omega) \,.
 \label{eq:G0aux}
\end{align}
Here, the inversion and multiplications are carried out for matrices in the site indices. The hybridization function is given in terms of the elements with impurity index $f$
\begin{align}
 \DRaux &= 1/g^R_0(\omega) - 1/G_{0ff}^R(\omega) \,,\nonumber\\
 \DKaux &= G_{0ff}^K(\omega)\,/\,|G_{0ff}^R(\omega)|^2 \,.
 \label{eq:Daux}
\end{align}
A single evaluation of the hybridization function is at most of $\mathcal{O}({N_B}^3)$ and thus not time consuming. However, for a large number of bath parameters ($\gtrsim20$) the multidimensional optimization problem may become demanding and appropriate methods are needed. In particular, a parallel tempering approach has proven to be effective, which is discussed in some more detail in \app\ref{app:opt}.

Beyond the requirement $\Daux \approx \Dph$, complete freedom exists in choosing a suitable auxiliary system. For the many-body solution with MPS it is convenient to allow for nearest neighbor terms in the Lindblad couplings only, i.e. to restrict not only $\vv{E}$ but also the matrices $\vv{\ga1}$ and $\vv{\ga2}$ to a tridiagonal form. In this way one ends up with a geometry where the impurity couples to a bath with n.n. terms only. As discussed below, the bipartite entanglement entropy of \aim can be reduced when imposing further that $\ga1_{i,j}$ has nonzero terms only for bath sites in one of the chains, e.g. for $i,j>f$, and $\ga2_{i,j}$ on the other side, i.e. for $i,j<f$. For the latter restriction we found that it affects the quality of the fit only in a minor way but significantly improves the applicability of MPS.

It is important to note that the relevant energy scale for the mapping procedure is not $\Gamma$ but the bandwidth $2D$. For a certain \aimns, one can adjust to different $\Gamma$-values by simply rescaling all terms in $\vv{E}$ with index $f$, i.e. the hoppings to the impurity site, without changing other properties of $\Daux$.~\cite{footnote_gamma} On the other hand, one can rescale the whole hybridization function by multiplying the matrices $\vv{E}$, $\vv{\ga1}$ and $\vv{\ga2}$ by the desired factor. Therefore, the complexity of the mapping procedure is dominated by the smallest $\omega$-scale compared to the largest one. For the flat band model, this essentially means that one has to regard $T$ and $\phi$ in units of $D$. With increasing number of bath sites $N_B$ we observe that sharper features can be resolved. Therefore, a maximal considered value of $N_B$ converts to a lower bound for the ratio of temperature $T$ to bandwidth $2D$ which can be reproduced by $\Daux$. More details on the 
mapping procedure are given in \app\ref{app:conv_NB} and \tcite{arri.15u}.

\subsection{Many-body solution}\label{ssec:manybody}

\subsubsection{Super-fermion representation}\label{sssec:superfermion}
As introduced in \tcites{dz.ko.11,schm.78} and made use of in \tcites{do.nu.14,ar.kn.13}, the Lindblad equation (\ref{eq:Leq}) can be recast into a standard operator problem when considering an augmented fermion Fock space with twice as many sites. We use the notation of \tcite{dz.ko.11}, to which we refer to for further details, in combination with a particle-hole transformation in the ``tilde'' space.~\cite{footnote_Ivac} The so-called left-vacuum reads
\begin{equation}
 \ket{I} = \sum_{\{n_{i\sigma}\}} (-i)^{N\left(\{n_{i\sigma}\}\right)} \ket{\{n_{i\sigma}\}} \otimes \ket{\widetilde{\{\bar{n}_{i\sigma}\}}}\,.
 \label{eq:Ivac}
\end{equation}
The summation runs over all possible many-body basis states $\ket{\{n_{i\sigma}\}}$ of the original system and $\ket{\widetilde{\{\bar{n}_{i\sigma}\}}}$ specifies those in the tilde system with inverted occupation numbers. $N\left(\{n_{i\sigma}\}\right) = \sum_{i\sigma}n_{i\sigma}$ is the total number of particles in state $\ket{\{n_{i\sigma}\}}$. 

The left-vacuum maps the density operator $\rho(t)$ onto the state vector $\ket{\rho(t)} = \rho(t) \ket{I}$. Thermodynamic expectation values are determined in this framework by expressions of the form $\braket{O(t)} = \bra{I}O\ket{\rho(t)}$. When evaluating $\left(\li \rho\right)\ket{I}$ for the Lindblad equation (\ref{eq:Leq}), one finds
\begin{equation}
\frac{d}{dt}\ket{\rho(t)} = L \ket{\rho(t)} \,,
\end{equation}
where the super-operator $\li$ is replaced by an ordinary non-Hermitian operator $L$. In vector notation
\begin{equation}
 \vv{c}_\sigma^\dagger = \left(c_{0\sigma}^\dagger,\dots,c_{N_B\sigma}^\dagger,\tilde c_{0\sigma}^\dagger,\dots,\tilde c_{N_B\sigma}^\dagger \right)\,,
\end{equation}
with $c_{i\sigma}^\dagger$/$c_{i\sigma}^\nag$ and $\tilde c_{i\sigma}^\dagger$/$\tilde c_{i\sigma}^\nag$ fermionic operators in the original and in the tilde system, respectively, the Lindblad operator $L$ is given by
\begin{align}
 iL &=  \sum_\sigma \vv{c}_\sigma^\dagger \begin{pmatrix}  \vv{E}+i\vv \Omega & 2 \vv \Gamma^{(2)}\\ -2 \vv \Gamma^{(1)} &  \vv E-i \vv \Omega \end{pmatrix}  \vv{c}_\sigma^\nag  - 2\,\text{Tr}\left(\vv E+i\vv \Lambda\right) \nonumber \\
 &\hspace{1em}  + U\left(n_{f\up} n_{f\dw} - \tilde{n}_{f\up} \tilde{n}_{f\dw} + \sum_\sigma \tilde{n}_{f\sigma} + 1  \right) \,,
 \label{eq:opL}
\end{align}
where $\vv\Omega = \vv{\ga2}-\vv{\ga1}$ and $\vv\Lambda= \vv{\ga1}+\vv{\ga2}$. Clearly, $L$ conserves the total particle number per spin $\sum_{i}\left(n_{i\sigma}+\tilde n_{i\sigma}\right)$. The steady state $\ket{\rho_\infty} = \lim\limits_{t\to\infty}\ket{\rho(t)}$ as well as $\ket{I}$ are situated in the half-filled, spin-symmetric sector. Steady state expectations values and correlation functions are calculated by~\cite{do.nu.14,footnote_G_AB}
\begin{equation}
\braket{A(t)B} = \bra{I}A e^{Lt}B\ket{\rho_\infty}\,,\hspace{1ex}\mathrm{for}\hspace{1ex}t\geq0\,.
\end{equation}

\begin{figure}
\begin{center}
\includegraphics[width=0.45\textwidth]{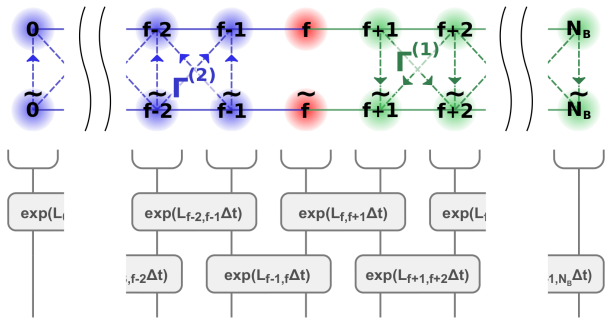}
\caption{(Color online) The upper part of the figure shows a schematic drawing of the auxiliary system in the super-fermion representation, for $N_B$ bath sites and with the impurity located at the central site $f = N_B/2$. The upper chain corresponds to original sites and the lower chain to the additionally introduced ``tilde'' sites, see \se\ref{sssec:superfermion}. In the chosen fit restriction, see \se\ref{ssec:mapping}, the coupling terms of the Lindblad operator $L$, \eq{eq:opL}, represent a ladder geometry with cross links. $\ga1$ causes a directional hopping from the upper to the lower chain and for $\ga2$ it is vice versa. Moreover, $\ga2_{i,j}$ is nonzero only for $i,j<f$ and $\ga1_{i,j}$ only for $i,j>f$. In the lower part, in grayscale, we schematically depict the tensor network for TEBD (\se\ref{sssec:mps}), where $L$ is decomposed in nearest neighbor terms $L_{i,i+1}$ which are applied in an alternating manner.}
\label{fig:IMaux_ladder}
\end{center}
\end{figure}

For tridiagonal matrices $\vv{E}$, $\vv{\ga1}$ and $\vv{\ga2}$, see \se\ref{ssec:mapping}, the coupling terms in \eq{eq:opL} represent a ladder system as depicted in \fig{fig:IMaux_ladder}. Sites on the original and the tilde system with the same index ($i=j$) or $i=j\pm1$ are coupled with rates $\ga1_{i,j}$ and $\ga2_{i,j}$ by a directional hopping. The restriction of $\ga2_{i,j}$ to the left side $i,j<f$ and for $\ga1_{i,j}$ to the right side $i,j>f$ leads to the situation that a circular current flows through the system. In this geometry one finds the tendency that sites on the left are filled in the original system and empty in the tilde system, whereas for the right side it is vice versa. This limits the possible hopping processes inside the chains and is in favor of a small bipartite entanglement entropy.~\cite{wo.mc.14}

\subsubsection{Matrix product states}\label{sssec:mps}
A large amount of literature exists on MPS in general and for Lindblad-type problems in particular.~\cite{scho.11,vida.04,wh.fe.04,zw.vi.04,ba.sc.09,ka.ba.12,wo.mc.14,mccu.07,weic.12,pr.zn.12,pr.zn.09,bo.ch.14,bo.la.14u,cu.ci.15,ma.fl.15u,ve.ga.04,we.ja.14u} Here we briefly state the governing equations for the well-known MPS methods made use of in this work.

We combine sites with the same index $i$ in the original and in the tilde system to one ``MPS-site'', with a local Hilbert space dimension $d=16$ (see also \fig{fig:IMaux_ladder}). For the resulting one-dimensional chain of sites it is straight-forward to write down a MPS representation:~\cite{scho.11}
\begin{equation}
  \ket{\rho} \,=\, \sum\limits_{\{s_i\}}c_{\{s_i\}}\ket{\{s_i\}}  \,=\,  \prod_{i=0}^{N_B} \left( \sum\limits_{s_i=1}^d \vv{A}^{s_i}_i \right)\ket{\{s_i\}}\,.
  \label{eq:rho_mps}
\end{equation}
Here, $\ket{\rho}$ is a generic many-body state with coefficients $c_{\{s_i\}}$ and $\vv{A}^{s_i}_i$ represents MPS matrices for site $i$ with local quantum numbers $s_i$.~\cite{footnote_mps}
The mapping \eq{eq:rho_mps} is exact for matrices which are exponentially large in $N_B$. However, even for much smaller matrix dimensions $\chi \ll d^{N_B/2}$ a very accurate representation of $\ket{\rho}$ is possible in many cases. For the auxiliary systems considered in this work, see also \se\ref{ssec:entanglement}, $\chi\approx1000$ is sufficient when making use of Abelian symmetries of the Lindblad operator \eq{eq:opL}. Concerning the positivity of $\rho$, one should note that the form of \eq{eq:rho_mps} does not ensure it per construction.\cite{ve.ga.04,we.ja.14u} However, we did not encounter unphysical results even for very small values of $\chi$.

In order to calculate observables, a MPS representation of $\ket{I}$ is needed. One finds that \eq{eq:Ivac} can be recast into a state with $\chi=1$, i.e. a product state, in which $\ket{I}$ is maximally entangled between original and tilde sites for the same index $i$. This is analogous to a purification of the identity operator.~\cite{zw.vi.04,ba.sc.09,pr.zn.12}

When rewriting the Lindblad operator \eq{eq:opL} with tridiagonal matrices $\vv{E}$, $\vv{\ga1}$ and $\vv{\ga2}$ in form of a matrix product operator, one has couplings of n.n. sites only. This enables us to use very efficient time evolution techniques as for example the time evolving block decimation (TEBD).~\cite{vida.04} Here, a Trotter decomposition is used to split the full time evolution $\exp(L\Delta t)$ into small parts $\exp(L_{i,i+1}\Delta t)$ for neighboring sites, and terms with even and odd $i$ are applied in an alternating manner, see also \fig{fig:IMaux_ladder}. In this work we use splitting methods accurate to second order in $\Delta t$.~\cite{footnote_splitting,stra.68,mc.at.92,wa.ma.12} We found that reducing the time step to $\Delta t = 0.01\,\Gamma^{-1}$ for the steady state and to $\Delta t = 0.05\,\Gamma^{-1}$ for the Green's functions is usually sufficient.

To obtain the desired steady state correlation functions of \aimns, for example $G^<$, we proceed as follows:
\begin{enumerate}
 \item Calculate the steady state $\ket{\rho_\infty}$ by time evolution with TEBD. Successively smaller time steps $\Delta t$ are used in order to eliminate the Trotter error. Static observables and $L\ket{\rho_\infty} = 0$ may serve as convergence criterion.~\cite{footnote_timeevo}
 \item Apply $c^\nag_{f\sigma}$ to $\ket{\rho_\infty}$ and time evolve the excited state to get $G_{\sigma}^<(t_n) = i\bra{I}c_{f\sigma}^\dagger e^{Lt_n} c^\nag_{f\sigma}\ket{\rho_\infty}$ at discrete points in the time domain.
 \item Employ linear prediction on the data $G^<(t_n)$ and thereafter a Fourier transformation to obtain $G^<(\omega)$ in the frequency domain.~\cite{ba.sc.09,wh.fe.04,footnote_G_AB,footnote_G_phsymm}
\end{enumerate}

\section{Results}\label{sec:results}
\begin{figure}
\begin{center}
\includegraphics[width=0.45\textwidth]{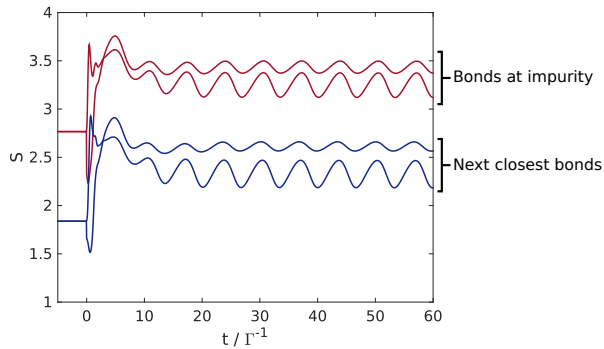}
\caption{(Color online) Temporal evolution of the bipartite entanglement entropy $S$ in a typical \aimns with $N_B=12$, representing the case $\phi = 1\,\Gamma$. The system is in the steady state for $t<0$ and $c_{f\sigma}$ is applied to $\ket{\rho_\infty}$ at $t=0$. We show $S(t)$ where it is largest, namely for the innermost bonds at the impurity, as well as for the next ones to the outside.} 
\label{fig:Ent_t}
\end{center}
\end{figure}
\begin{figure}
\begin{center}
\includegraphics[width=0.4\textwidth]{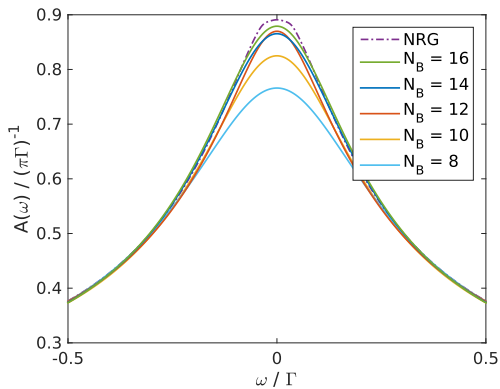}
\caption{(Color online) Spectral function in equilibrium: plotted for different number of bath sites $N_B$ and compared with reference data from NRG.~\cite{footnote_nrg} Results are for $U=6\,\Gamma$ and flat band leads, \eq{eq:gR_lambda}, with $D=10\,\Gamma$ and $T = 0.05\,\Gamma$.} 
\label{fig:Aw_phi0}
\end{center}
\end{figure}
Before focusing on the nonequilibrium physics of the single impurity Anderson model, we briefly discuss the bipartite entanglement entropy of the auxiliary impurity model, and a benchmark for the equilibrium case. After that the spectral properties as a function of bias voltage are presented for two different temperatures, one clearly below and one above the Kondo temperature $T_K$. Furthermore, the bias dependence of observables such as the current and the double occupancy is discussed. In the last part of this paper a different density of states in the leads is considered, which allows to better resolve the physics at low temperatures and low bias voltages.

\subsection{Entanglement scaling}\label{ssec:entanglement}
Matrix product states are an efficient representation of many-body states with a low bipartite entanglement entropy $S$. The required matrix dimension $\chi$ at a certain bond $(i,i+1)$ scales exponentially with the entropy at this bond, $S_{i,i+1}$. From Hermitian systems it is known, that ground states of gapped, one-dimensional systems obey an area law and are thus well-suited for MPS. Also an evolution in imaginary time converges well, but the real time evolution of excited states may become problematic due to a build-up of entanglement.~\cite{scho.11} For the auxiliary impurity model investigated here, the behavior appears to be opposite. In general, the steady state $\ket{\rho_\infty}$ of \aim does not fulfill an area law and instead an increase of $\max_iS_{i,i+1}$ with increasing system size $N_B$ is observed.~\cite{bo.la.14u} Despite this, the time evolution of excited states is unproblematic, likely because of the damping involved, and the long time limit can easily be reached.

We observe that the optimized parameters in \aim strongly depend on the number of bath sites and on the external, physical parameters ($\phi$, $T$, \dots).
Therefore, it is difficult to infer a reliable quantitative entanglement scaling with $N_{B}$.
Qualitatively we find that $\max_iS_{i,i+1}$ increases moderately with $N_B$ and slower than linear. The magnitude of the entanglement is considerably reduced by the restricted setup for \aim described in \se\ref{ssec:mapping}, which has the tendency towards a filled and an empty bath chain in the steady state. In this setup, $S_{i,i+1}$ takes on the largest value at the central bonds which connect to the impurity site and falls off quickly with distance from the center.

Independent of the actual scaling, the increase of bipartite entanglement with $N_B$ has the consequence that one is limited to certain system sizes. In this work we consider up to $N_B = 16$ with $\chi=1000$, which is feasible in a reasonable amount of time. Most likely, one would need higher values of $\chi$ in order to treat even larger systems precisely. We checked the reliability of the results presented below by increasing the matrix dimension to $\chi=1500$ in several cases, for different values of $\phi$ and $T$. Furthermore, the time evolution for the Green's functions was validated by reducing the time step to $\Delta t = 0.01\,\Gamma^{-1}$. Overall, we found in the worst cases relative differences in $A(\omega)$ up to $\mathcal{O}(10^{-3})$. These errors are small enough for our purposes, so that we focus in the following on the accuracy of the mapping procedure, i.e. versus $N_B$.

\begin{figure*}
\begin{center}
\includegraphics[width=0.4\textwidth]{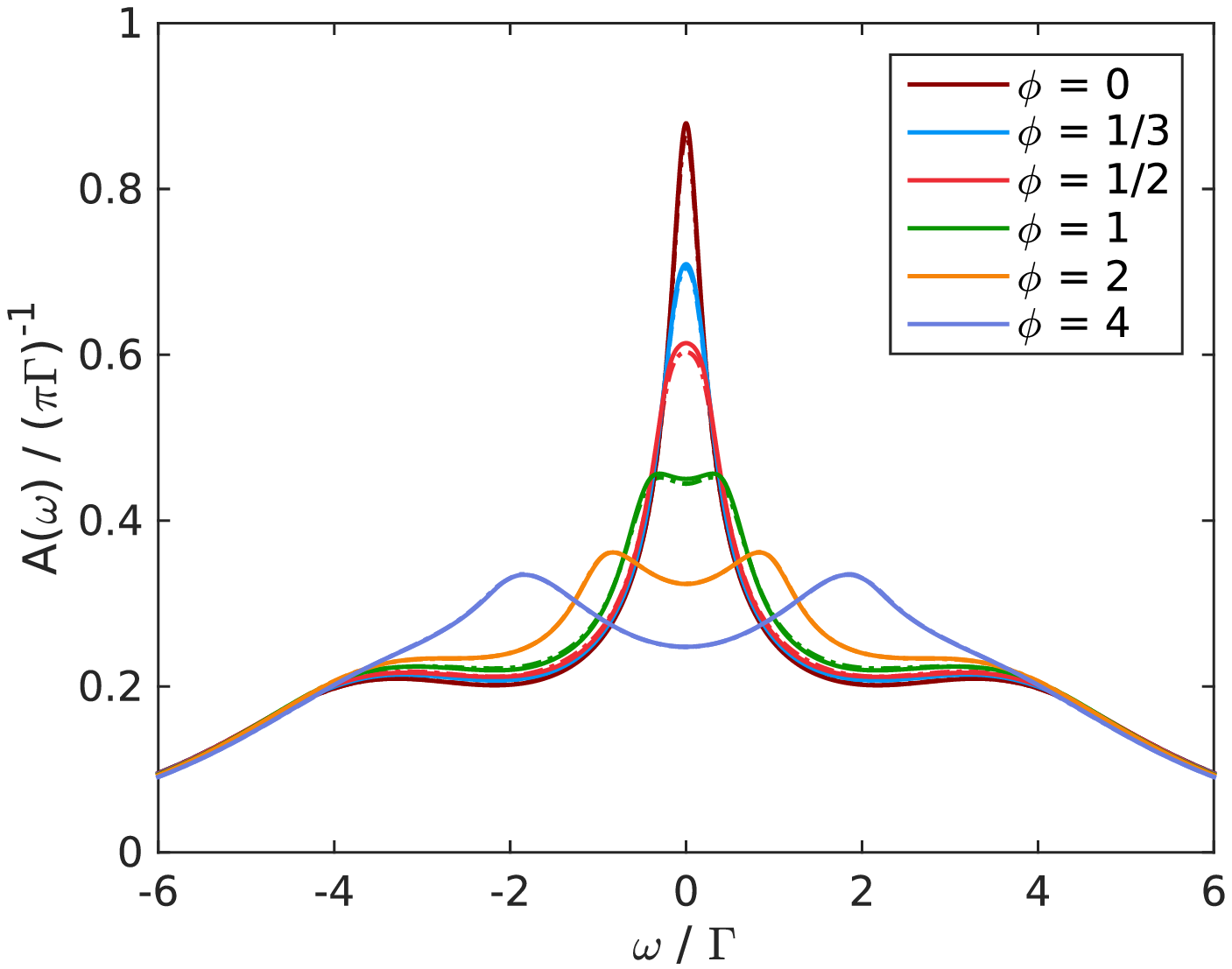}\hspace{1em}
\includegraphics[width=0.4\textwidth]{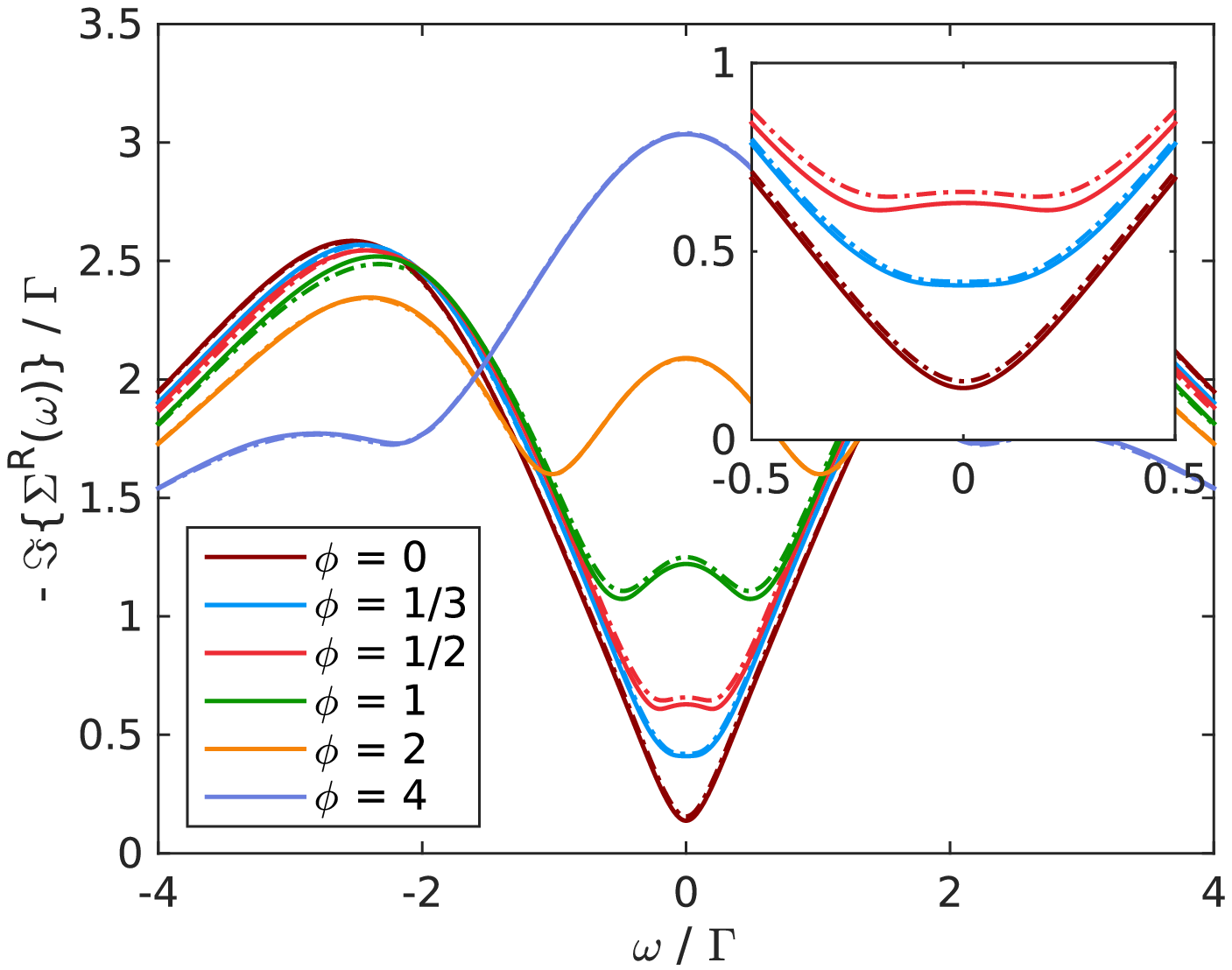}\vspace{-1ex}
\caption{(Color online) Bias-dependent spectral function (left) and retarded self-energy (right) for $T = 0.05\,\Gamma$. Solid lines correspond to calculations with $N_B = 16$ and dash-dotted lines to $N_B = 14$, but in many cases they cannot be distinguished. Bias voltage $\phi$ is given in units of $\Gamma$. Results are for $U=6\,\Gamma$ and flat band leads, \eq{eq:gR_lambda}, with $D=10\,\Gamma$.} \vspace{-1.5em}
\label{fig:Aw_SigR_phi_T0_05}
\end{center}
\end{figure*}
\begin{figure*}
\begin{center}
\includegraphics[width=0.4\textwidth]{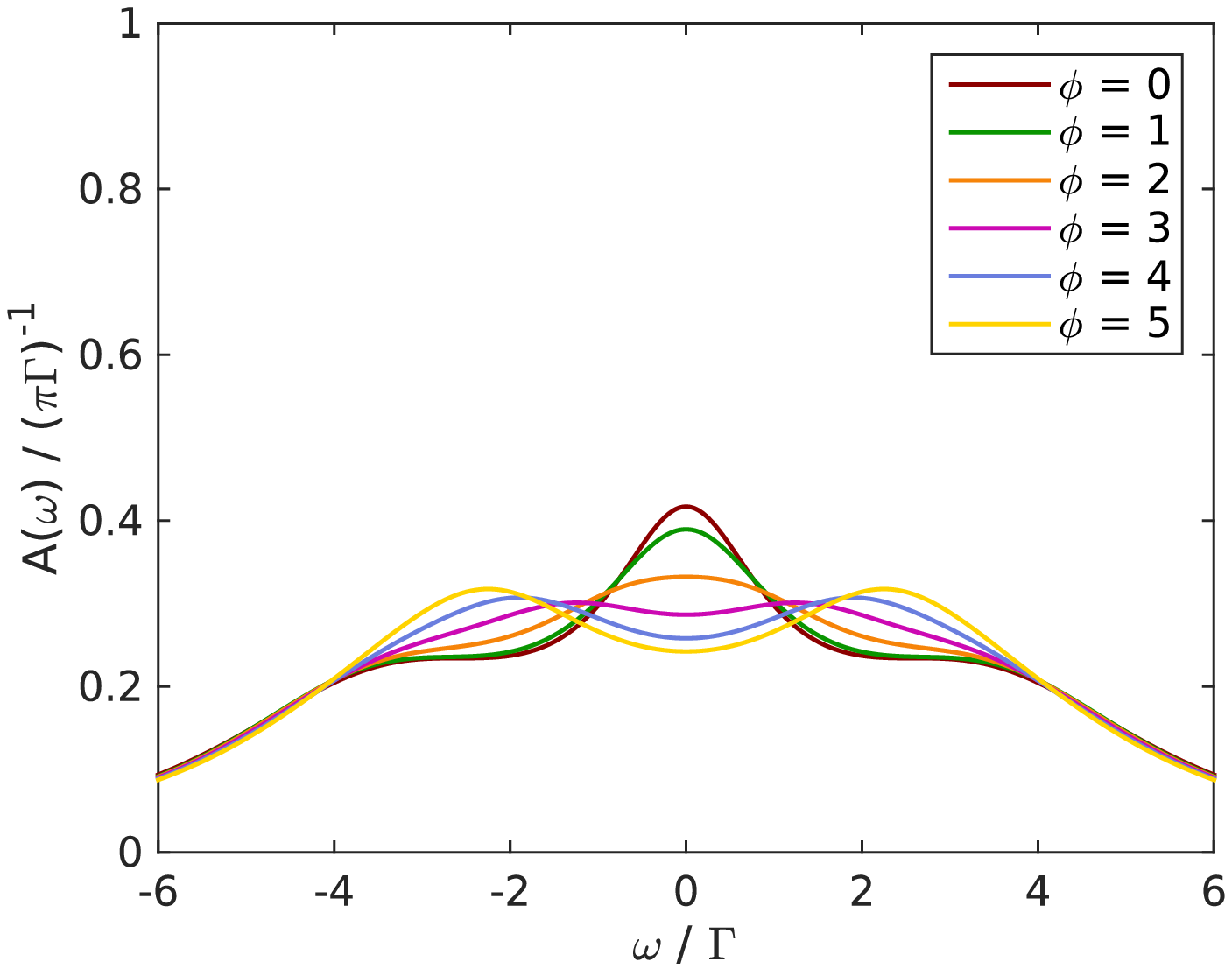}\hspace{1em}
\includegraphics[width=0.4\textwidth]{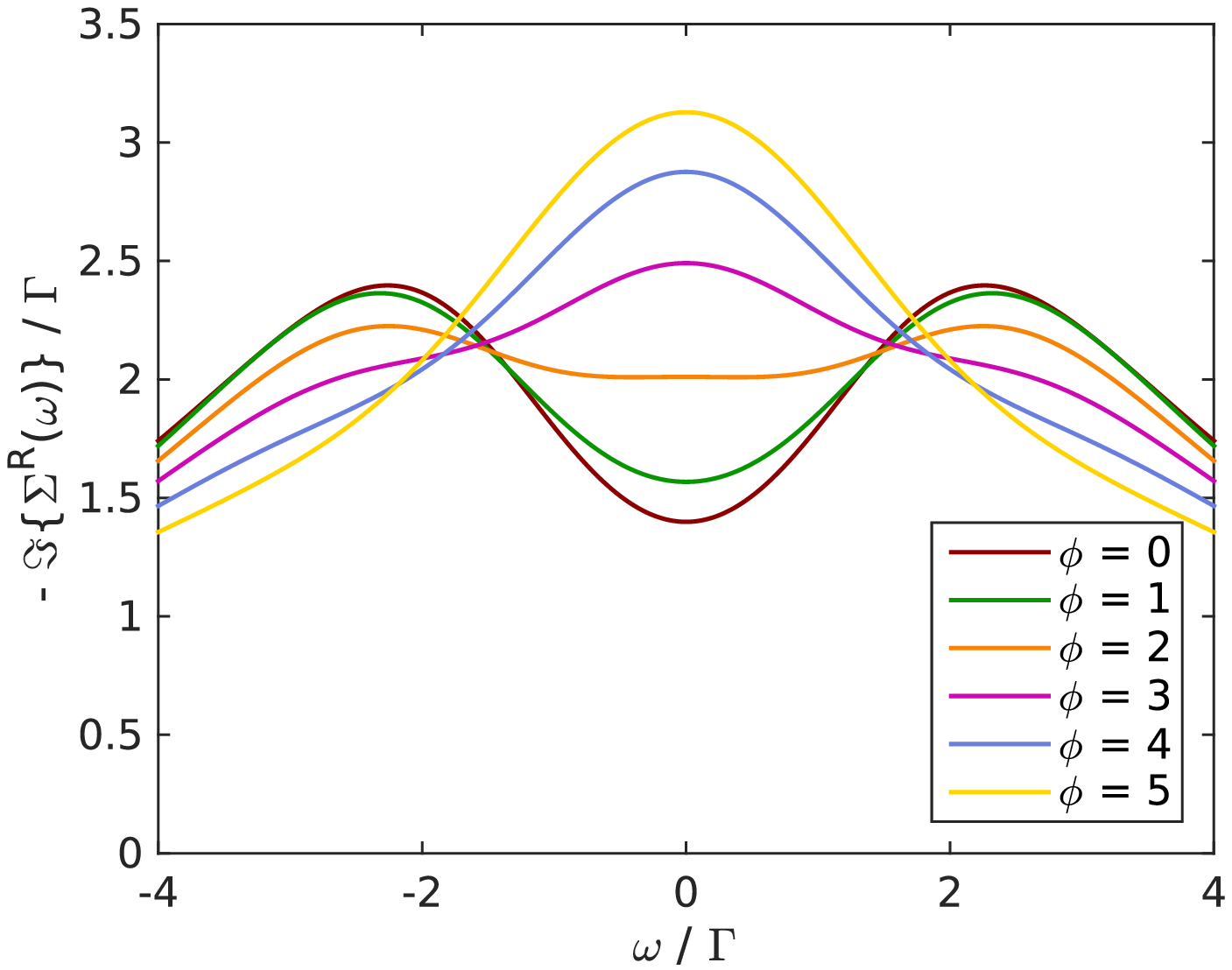}\vspace{-1ex}
\caption{(Color online) Bias-dependent spectral function (left) and retarded self-energy (right) for $T = 0.5\,\Gamma$. Calculations are performed with $N_B = 10$ and other parameters are the same as in \fig{fig:Aw_SigR_phi_T0_05}.} 
\label{fig:Aw_SigR_phi_T0_5}
\end{center}
\end{figure*}
\begin{figure*}
\begin{center}
\includegraphics[width=0.4\textwidth]{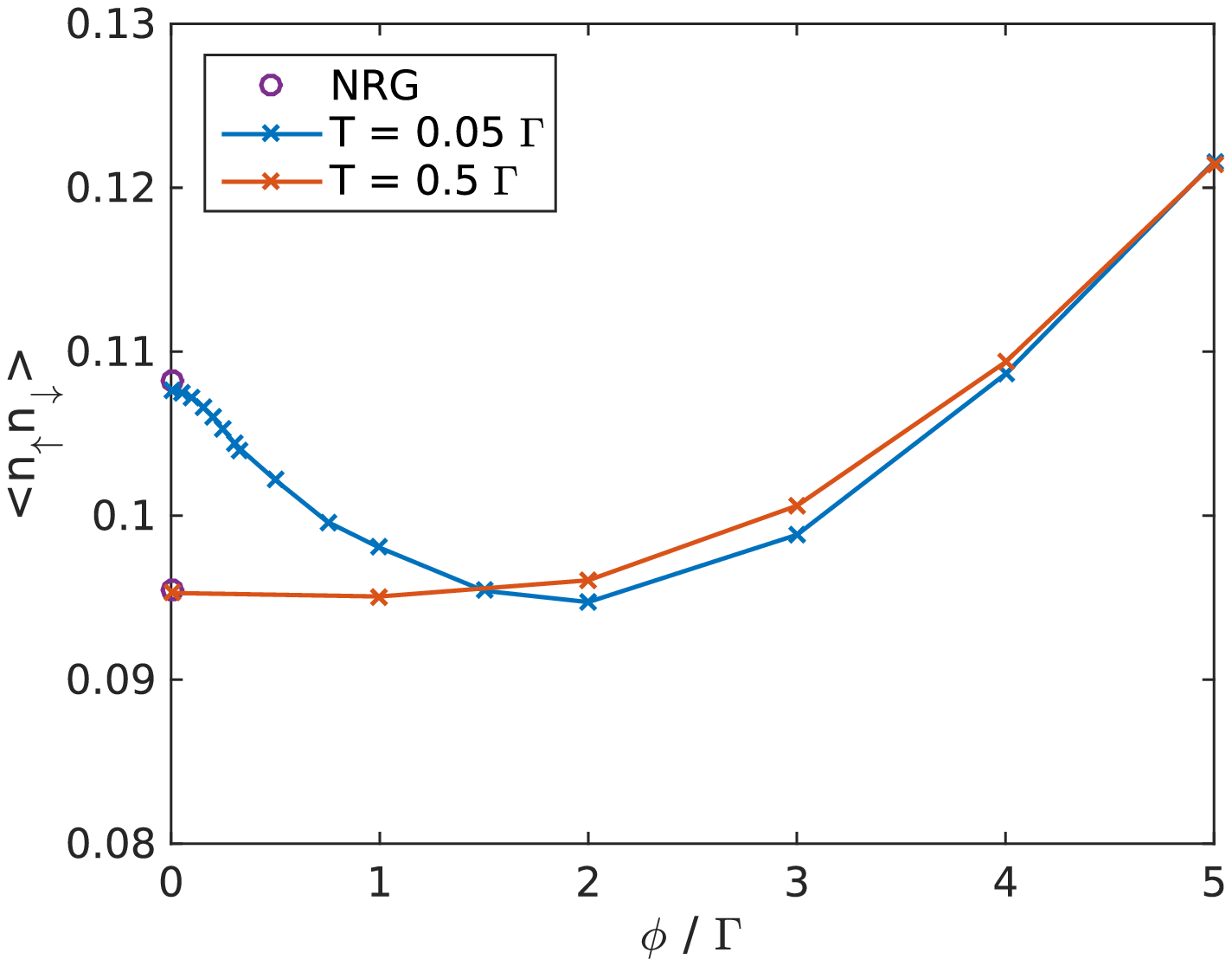}\hspace{1em}
\includegraphics[width=0.4\textwidth]{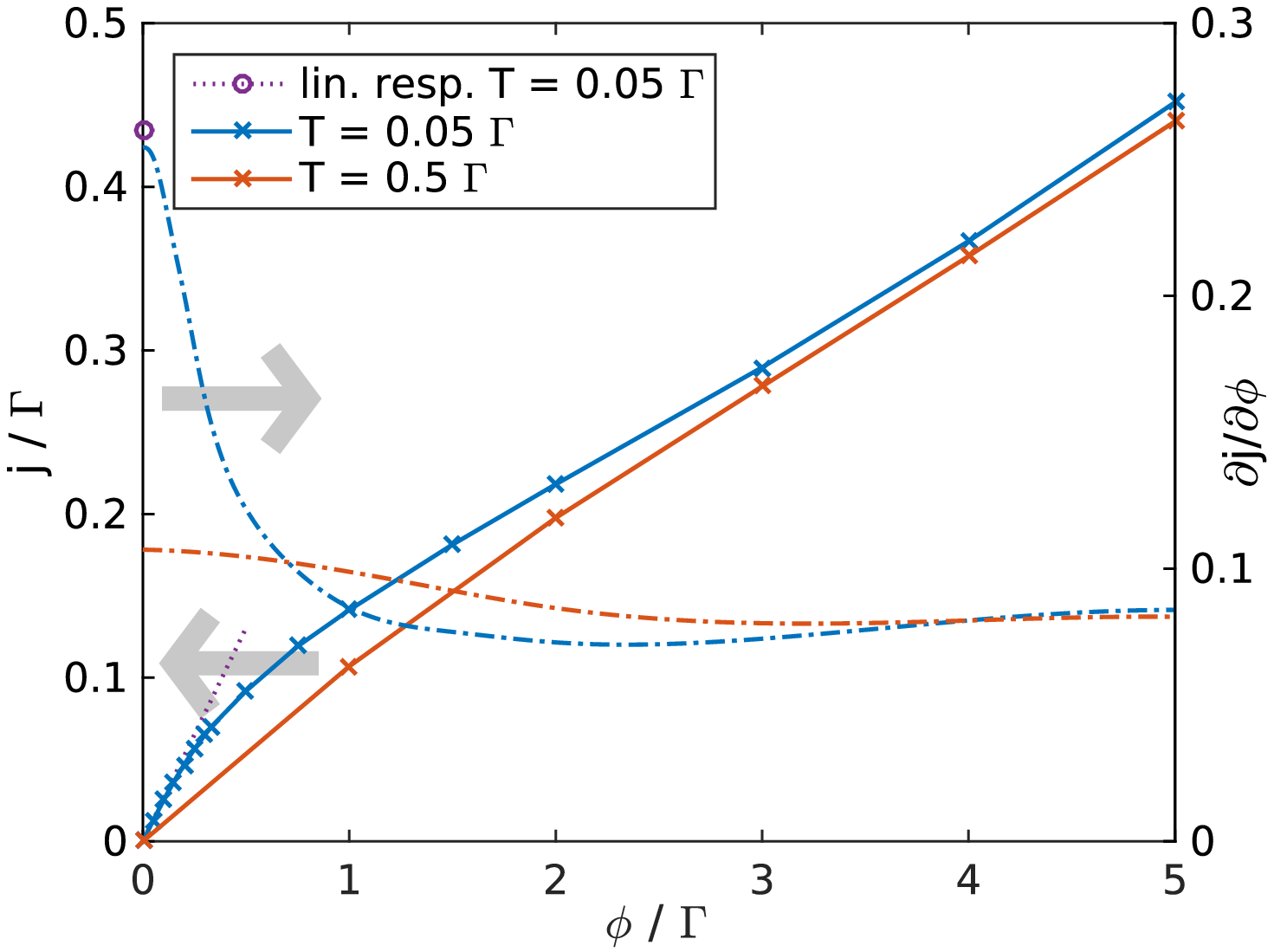}\vspace{-1ex}
\caption{(Color online) Double occupancy (left) and transport properties (right) as a function of bias voltage $\phi$. Current $j$ is depicted with solid lines and the differential conductance $\partial j/\partial\phi$ with dash-dotted lines. The latter is calculated with three-point Lagrange polynomials, based on the data for $j$ as marked in the plot. Results are shown for $T = 0.5\,\Gamma$ with $N_B = 10$, and for $T = 0.05\,\Gamma$ with $N_B = 14$. Other parameters are as in \fig{fig:Aw_SigR_phi_T0_05}. The linear response and equilibrium values for $\braket{n_{f\up} n_{f\dw}}$ are from NRG.~\cite{footnote_nrg} }
\label{fig:dd_j_phi}
\end{center}
\end{figure*}

To analyze the temporal evolution of $S$, a typical time-dependent case is shown in \fig{fig:Ent_t}. Here, $t<0$ indicates the steady state regime and at $t=0$ an annihilation operator is applied to $\ket{\rho_\infty}$ in order to calculate the lesser Green's function. To estimate the relevant time scale, $\iim\{G^<(t)\}$ (not plotted) drops from $0.5$ at $t=0$ to $10^{-2}$ at $t = 3\,\Gamma^{-1}$. As one can see, $S_{i,i+1}$ changes at first rapidly but saturates then and oscillates in time around a constant value. Thus, it is unproblematic to resolve $G^<(t)$ even on very large time scales. This was furthermore checked for small $N_B$ with an exact diagonalization solution as reference.~\cite{footnote_EDcheck} When inspecting the short-time behavior of $S_{i,i+1}$, an asymmetry is evident. This results from the application of an operator to the original system alone, without changing the tilde system.

\subsection{Spectral and transport properties}\label{ssec:results_flatband}
Before focusing on the nonequilibrium physics, we briefly present results for the equilibrium situation $\phi=0$ in \fig{fig:Aw_phi0}. Here, a quasi-exact solution is provided by means of NRG.~\cite{footnote_nrg} For $T=0.05\,\Gamma\approx T_K/4$ the system is well inside the Kondo regime and the peak height of $A(0) \pi \Gamma\approx 0.9$ almost fulfills the $T=0$ Friedel sum rule ($A(0) \pi \Gamma=1$ for $T\to 0$).~\cite{lang.66,la.am.61,footnote_nrgTk} Results obtained with AMEA are shown for different system sizes, with particular focus on the low energy physics. Noticeable differences are apparent for $N_B=8$, but, upon increasing the number of bath sites quick convergence is observed and excellent agreement with the NRG data is found. This shows that AMEA, especially with MPS, is a very accurate impurity solver also in the equilibrium case for $T>0$. 

Regarding the accuracy of the calculations, we can state that $N_B=12$ is essentially sufficient to provide reliable spectral data in equilibrium for $T = 0.05\,\Gamma$. However, for the nonequilibrium situations considered in the following one has to take into account that the accuracy of the mapping procedure is to some degree dependent on $\phi$. This is analyzed in detail in \app\ref{app:conv_NB} and here we solely want to note that the low bias $\phi\leq\frac{1}{3}\,\Gamma$ as well as the higher bias regime $\phi\geq2\,\Gamma$ converge more rapidly than the intermediate values, for $T = 0.05\,\Gamma$. For the larger values of $T$ used below the calculations are even easier, as one can achieve a very good mapping $\Daux\approx\Dph$ already for less than $N_B= 12$.

After this benchmark, we now study the steady state nonequilibrium spectral properties for two different temperatures, one below and one above the Kondo temperature. In \fig{fig:Aw_SigR_phi_T0_05} results are presented for $T = 0.05\,\Gamma$ and in \fig{fig:Aw_SigR_phi_T0_5} for $T = 0.5\,\Gamma$. In the first case, it is apparent that small bias voltages $\phi<\Gamma$ cause a decrease and smearing of the Kondo peak, whereas larger voltages result in a splitting, see also \tcites{wi.me.94,le.sc.01,ro.kr.01,ko.sc.96,fu.ue.03,nu.he.12,ande.08,ha.he.07,co.gu.14}. It is known that with increasing current, resonant spin-flip scattering is prevented due to decoherence. Despite this, distinct excitations are clearly visible even at rather high bias voltages and located approximately at the positions of the chemical potentials $\mu_{L/R} = \pm\frac{\phi}{2}$. This can be attributed to intra-lead processes, which however are strongly suppressed. In \fig{fig:Aw_SigR_phi_T0_05} we present furthermore 
the retarded self-energy, which enables us to better locate at which $\phi$-value splitting sets in. An upper bound can be estimated by the value $\phi = 0.5\,\Gamma$, where $-\iim\{\Sigma^R(\omega)\}$ exhibits two minima. In \se\ref{ssec:results_Lorentzian} we resolve the physics at low bias in some more detail.

In \fig{fig:Aw_SigR_phi_T0_5} the same system is considered for $T = 0.5\,\Gamma$. As expected, the features are much broader and the Kondo peak for $\phi = 0$ is strongly suppressed.~\cite{cost.00} Despite of this, one can still note splitting and weak excitations at $\mu_{L/R} = \pm\frac{\phi}{2}$ at rather high voltages $\phi \geq 3\,\Gamma$. In $-\iim\{\Sigma^R(\omega)\}$, only the result for $\phi = 2\,\Gamma$ exhibits two slight minima. One can thus infer, that the temperature dominates the decoherence processes on the impurity in this case and excitations at $\mu_{L/R}$ are further suppressed and strongly smeared out.

For both temperatures $T = 0.05\,\Gamma$ and $T = 0.5\,\Gamma$, we present two observables of interest, the double occupancy and the current, in \fig{fig:dd_j_phi}. The latter is obtained from the standard Meir-Wingreen expression.~\cite{me.wi.92,ha.ja,jauh} In the current it is obvious that the temperature strongly influences the low bias regime, as is expected from linear response considerations. Especially the differential conductance enables to resolve the low-bias physics and we find a typical Kondo behavior.~\cite{pl.sc.12,re.pl.14} At higher voltages $\phi \gtrsim 2\,\Gamma$, however, one observes for $T = 0.05\,\Gamma$ a slight increase of $\partial j/\partial\phi$ due to charge fluctuations. At even higher voltages ($\phi \gtrsim 3\Gamma$) both temperatures result in a similar linear current-voltage characteristic since the two spectral functions nearly merge into each other. The double occupancy $\braket{n_{f\up} n_{f\dw}}$ exhibits an interesting behavior for $T=0.05\;\Gamma<T_{K}$.
In this case, $\braket{n_{f\up} n_{f\dw}}$ and thus the charge fluctuation as well, exhibits a minimum (at $\phi \approx 2\,\Gamma$). It originates from two competing mechanisms evolving with increasing $\phi$: On the one hand, the enlarged transport window, approximately given by the interval $(-\frac{\phi}{2},\,\frac{\phi}{2})$, increases $\braket{n_{f\up} n_{f\dw}}$ and on the other hand, the suppression of resonant spin-flip scattering has the opposite effect. Apparently, the latter dominates initially at low bias. We observe a  similar behavior in the temperature dependence of the double occupancy $\braket{n_{f\up} n_{f\dw}}_{T}$ in the equilibrium case. We find a minimum in $\braket{n_{f\up} n_{f\dw}}_{T}$ at $T \approx 0.5\,\Gamma$. Therefore, the impurity is at this value in the local moment regime. When applying a bias voltage in the case of $T = 0.5\,\Gamma$, the double occupancy increases monotonically with $\phi$, as can be seen in \fig{fig:dd_j_phi}. One can therefore conclude that the Kondo 
effect and its suppression with increasing $\phi$ has a significant effect on the double occupancy. However, the particular position of the minimum is not related to $T_K$ but essentially only determined by the energy scale $\Gamma$, as discussed in \tcites{di.sc.13,ha.di.12}. The reliability of our results is corroborated by the close agreement for $\phi=0$ with the equilibrium values obtained by NRG (marked by circles in \fig{fig:dd_j_phi}).

\subsection{Low bias spectrum}\label{ssec:results_Lorentzian}
\begin{figure*}
\begin{center}
\includegraphics[width=0.4\textwidth]{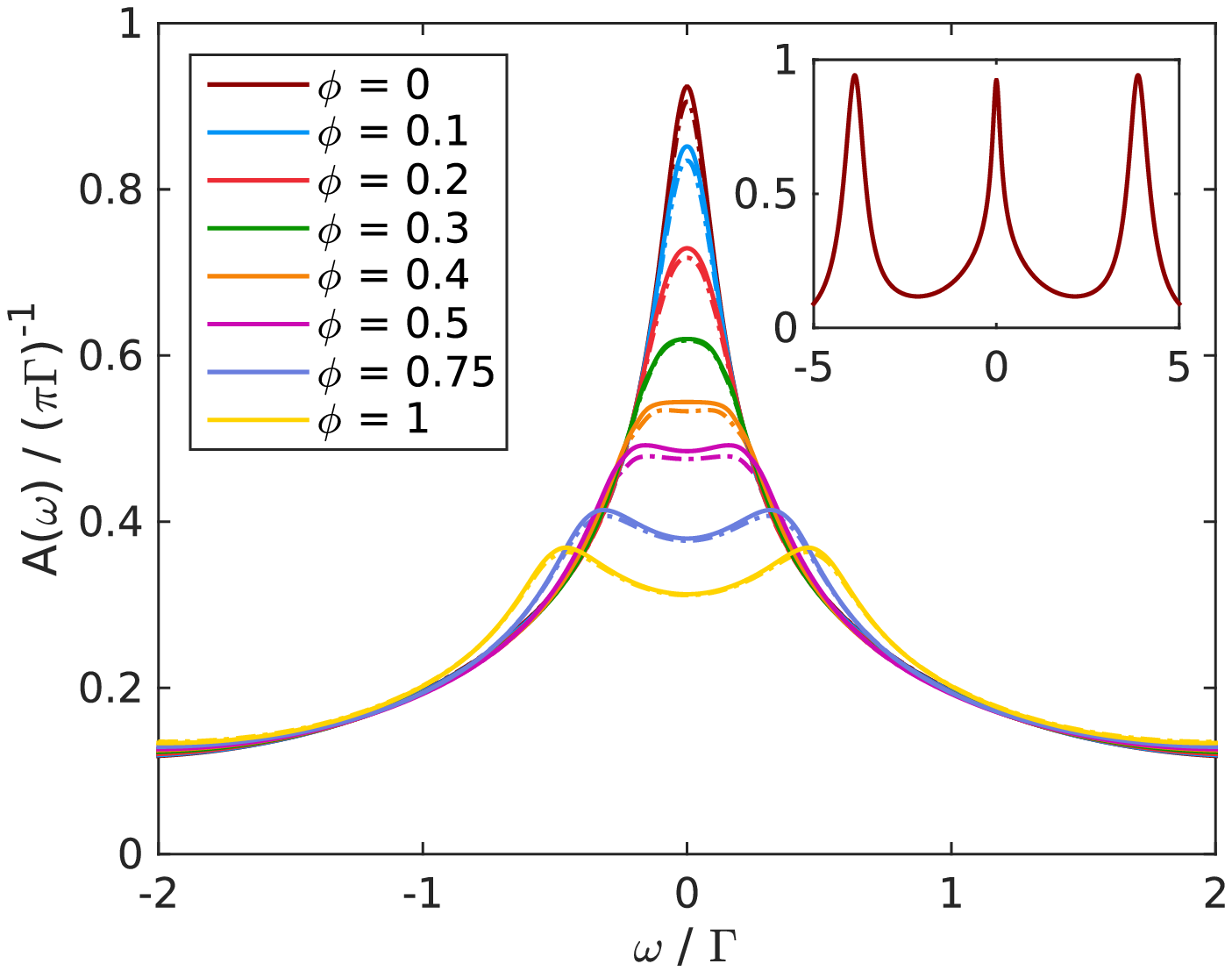}\hspace{1em}
\includegraphics[width=0.4\textwidth]{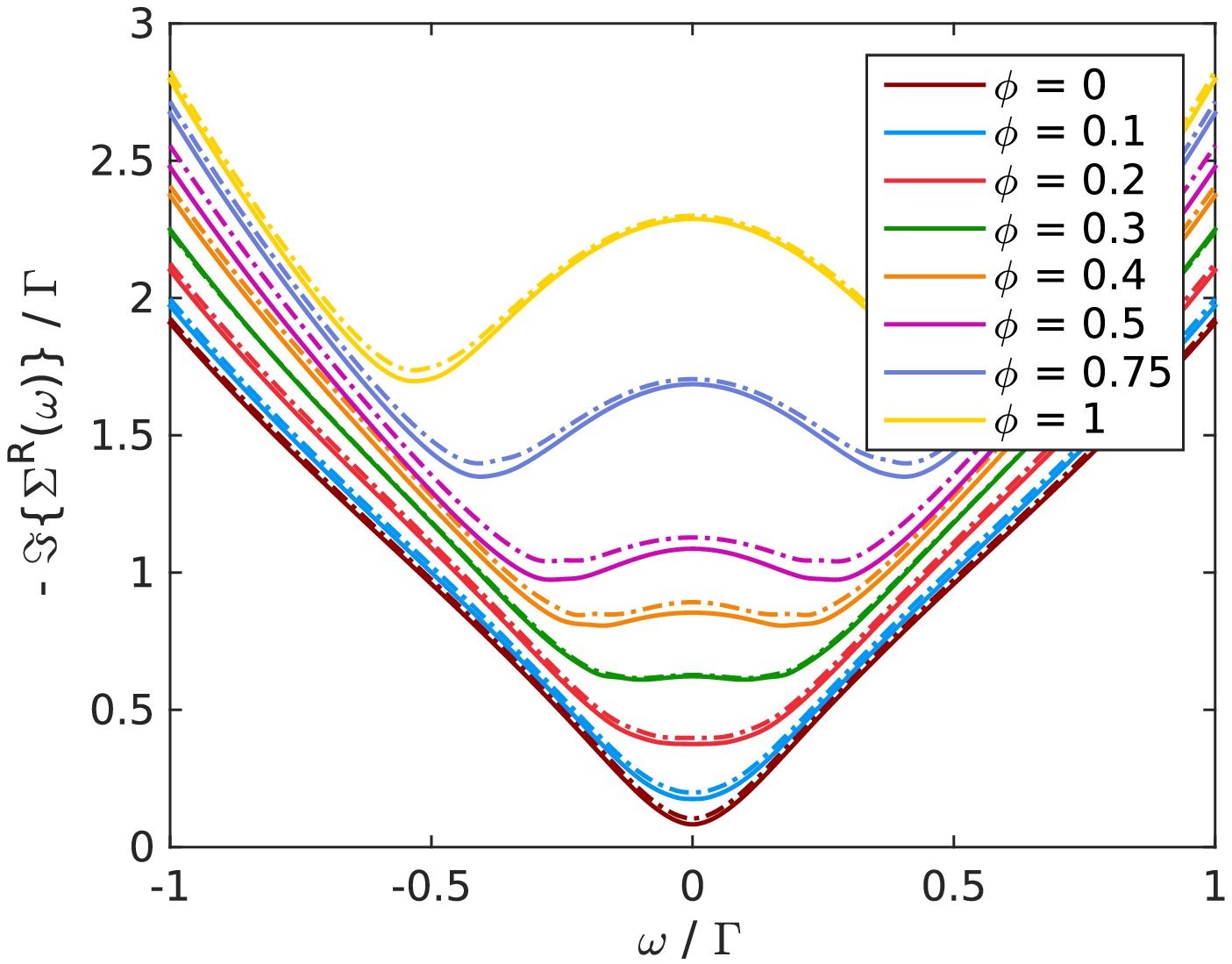}
\caption{(Color online) Bias-dependent spectral function (left) and retarded self-energy (right) for $U=6\,\Gamma$, $T = 0.02\,\Gamma$ and a Lorentzian density of states in the leads, see \eq{eq:gR_lambda_Lorentzian}. Solid lines correspond to calculations with $N_B = 16$ whereas dash-dotted lines to ones with $N_B = 14$. The bias voltage $\phi$ is in units of $\Gamma$. The inset on the left is for $N_B = 16$ and $\phi = 0$.} 
\label{fig:Aw_SigR_phi_Lorentzian}
\end{center}
\end{figure*}
In order to better resolve the low energy spectral properties of the Anderson impurity model, we now consider briefly the case of a Lorentzian density of states in the leads. In particular we replace \eq{eq:gR_lambda} with
\begin{equation}
 g^R_\lambda(\omega) =  \left( \omega -\varepsilon_\lambda + i \gamma \right)^{-1}\,,
 \label{eq:gR_lambda_Lorentzian}
\end{equation}
where $\varepsilon_{L/R} = \pm\frac{\phi}{2}$. This can be produced by a bath consisting of one site with on-site energy $\varepsilon_\lambda$ which further connects to a wide band given by $\gamma$. We take $\Gamma = -\iim\{\Delta^R(0)\}$ again as unit of energy and choose $\gamma = 5/\pi\,\Gamma$ together with $2{t'_\lambda}^2/\gamma = \Gamma$. The Keldysh component is given by \eq{eq:FDT} with $\mu_{L/R} = \pm\frac{\phi}{2}$, as before.

A Lorentzian density of states is particularly suited for AMEA, since the retarded part $\DRaux$ alone can be fitted exactly with a single bath site. This simplification does not apply to the Keldysh component with its Fermi edges. Still, one can expect that the mapping procedure is more accurate than for a flat density of states and indeed we find that we are able to reproduce $\Dph$ by $\Daux$ more precisely for the same $N_B$. As a result, we can reach lower $T$ with the same system sizes. For details on the achievable accuracy we refer to \app\ref{app:conv_NB}.

In particular we investigate the case $T = 0.02\,\Gamma$ and $U=6\,\Gamma$ and focus on bias voltages close to $T_K$. In addition to the lower temperature especially the smaller effective hybridization strength at the position of the Hubbard bands leads to an increased separation of Kondo and Hubbard features in the spectral function, and thus, to an improved resolution. This can be seen in the peaked structure of the inset in \fig{fig:Aw_SigR_phi_Lorentzian}. The smaller temperature allows us to analyze the behavior for lower bias voltages down to $\phi = 0.1\,\Gamma$.~\cite{footnote_Lorentzian_Tphi} Also for this setup we find a similar dependence of the spectrum as a function of voltage as before, only at a decreased energy scale. The self-energy in \fig{fig:Aw_SigR_phi_Lorentzian} indicates that a splitting is first perceptible at a bias of $\phi \approx 0.2-0.3\,\Gamma$. 
From our data we can thus conclude that for bias voltages just above the Kondo temperature, a clear splitting of the Kondo resonance into a simple two-peak structure occurs.

\section{Conclusions}\label{sec:conclusion}
In this work we presented an improved formulation of AMEA, introduced in \tcites{ar.kn.13,do.nu.14}, obtained by employing matrix product states for the solution of the auxiliary master equation in the interacting case. This allowed us to treat larger auxiliary systems with more optimization parameters for the mapping procedure, as compared to the ED based solver in \tcite{do.nu.14}. This is crucial, since the accuracy in AMEA increases exponentially with the number of optimization parameters. As a result, we obtained well-converged spectral data and static observables, whose accuracy for the equilibrium case were comparable to NRG down to low temperatures and for large interactions.
More specifically, in the calculations presented here, we were able to investigate the steady state properties of the single impurity Anderson model as a function of bias voltage $\phi$ and at temperatures $T$ well below the Kondo temperature $T_K$. In the spectral function we obtained a prominent Kondo peak for $\phi=0$ and $T\approx T_K/4$, which compared very well to an equilibrium NRG calculation, and a broadening and subsequent splitting of the peak when considering $\phi>0$. Also for the case of a Lorentzian density of states in the leads, which enabled us to lower the temperature to $T\approx T_K/10$, we found no evidence of a different behavior than a simple splitting of the Kondo peak. In order to locate the value of $\phi$ at which the peak starts to split, it was advantageous to inspect the retarded self energy. From this we concluded that two excitations become visible for bias voltages just above the Kondo temperature, at $\phi \approx 1-2\,T_K$.

For the many-body solution with MPS it was of advantage to adjust the geometry of the auxiliary system and possible modifications were discussed. As in other studies of Lindblad problems with MPS, we found an increase of the bipartite entanglement entropy $S$ with system size $N_B$.~\cite{bo.la.14u} However, the increase was moderate and slower than linear, which made it possible to treat auxiliary open systems up to $N_B\approx16$ reliably and within a rather short computation time (a couple of days). The value $N_B=16$ is by no means a ``hard limit" and much larger systems are expected to be feasible, especially when including additionally non-Abelian symmetries.~\cite{mccu.07,weic.12} 

In general, the present MPS extension of AMEA constitutes a versatile and very accurate impurity solver for both equilibrium and nonequilibrium steady state situations. Compared to the ED based solver presented in \tcite{do.nu.14}, the computation time is longer but the achievable accuracy is much higher.
Therefore, the MPS impurity solver is especially suited for situations in which a high spectral resolution is needed and a detailed investigation of the underlying physics is desired.

\begin{acknowledgments}
This work was supported by the Austrian Science Fund (FWF): P24081 and P26508, as well as SFB-ViCoM projects F04103 and F04104, and NaWi Graz. The calculations were partly performed on the D-cluster Graz and on the VSC-3 cluster Vienna. M.G. acknowledges support by the Simons Foundation (Many Electron Collaboration) and by Perimeter Institute for Theoretical Physics.
We are grateful to Frauke Schwarz, Jan von Delft and Andreas Weichselbaum, as well as Sabine Andergassen, Martin Nuss, Markus Aichhorn, Marko Znidaric, Ulrich Schollw\"ock and Valentin Zauner for fruitful discussion. Furthermore, the authors want to thank Rok {\ifmmode \check{Z}\else\v{Z}\fi{}}itko for providing his open source code NRG Ljubljana.\cite{footnote_nrg}
\end{acknowledgments}

\appendix
\begin{figure}
\begin{center}
\includegraphics[width=0.45\textwidth]{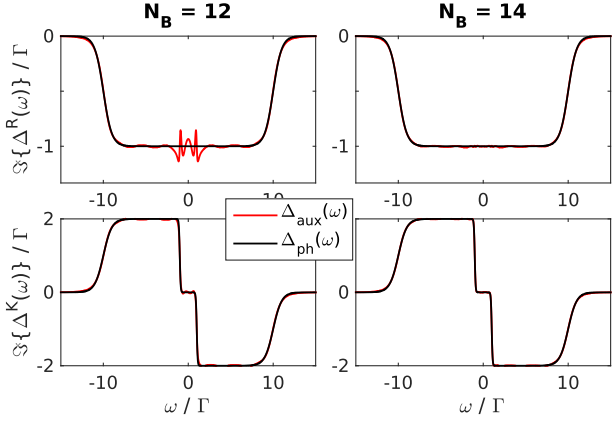}
\caption{(Color online) Hybridization function $\Daux$ as obtained from minimizing the cost function \eq{eq:costfunc} with $\omega_c = 15\,\Gamma$ and $W(\omega)=1$, for the flat band model \eqs{eq:Delta},\thinspace(\ref{eq:gR_lambda}) with $\phi=2\,\Gamma$ and $T = 0.05\,\Gamma$. Results on the left are for $N_B=12$ and on the right for $N_B=14$.}
\label{fig:Fit}
\end{center}
\end{figure}
\begin{figure*}
\begin{center}
\includegraphics[width=0.75\textwidth]{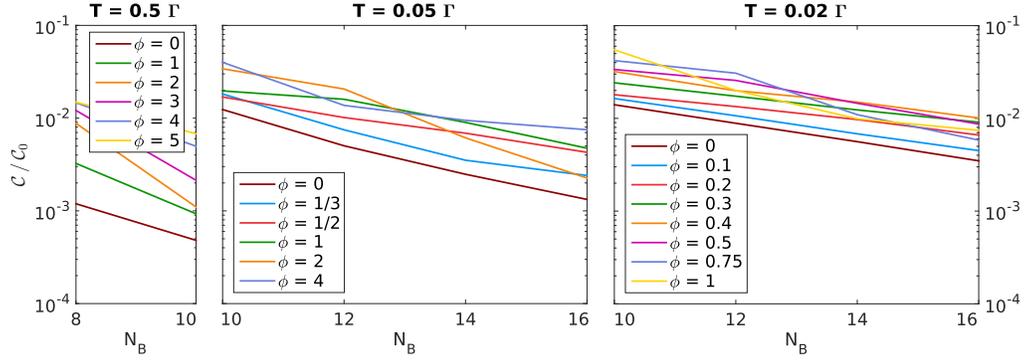}
\caption{(Color online) Convergence of $\Daux$ with increasing $N_B$. Results with $T = 0.5\,\Gamma$ and $T = 0.05\,\Gamma$ are for the flat band case \eqs{eq:Delta},\thinspace(\ref{eq:gR_lambda}), and the ones with $T = 0.02\,\Gamma$ are for the Lorentzian density of states \eq{eq:gR_lambda_Lorentzian}. For the cost function $\mathcal{C}$, \eq{eq:costfunc}, we chose $W(\omega)=1$ as well as $\omega_c = 15\,\Gamma$ for the flat band model and $\omega_c = 5\,\Gamma$ for the Lorentzian case. The normalization $\mathcal{C}_0$ refers to the value of $\mathcal{C}$ for $\Daux\equiv0$.}
\label{fig:conv_N}
\end{center}
\end{figure*}
\begin{figure*}
\begin{center}
\includegraphics[width=0.97\textwidth]{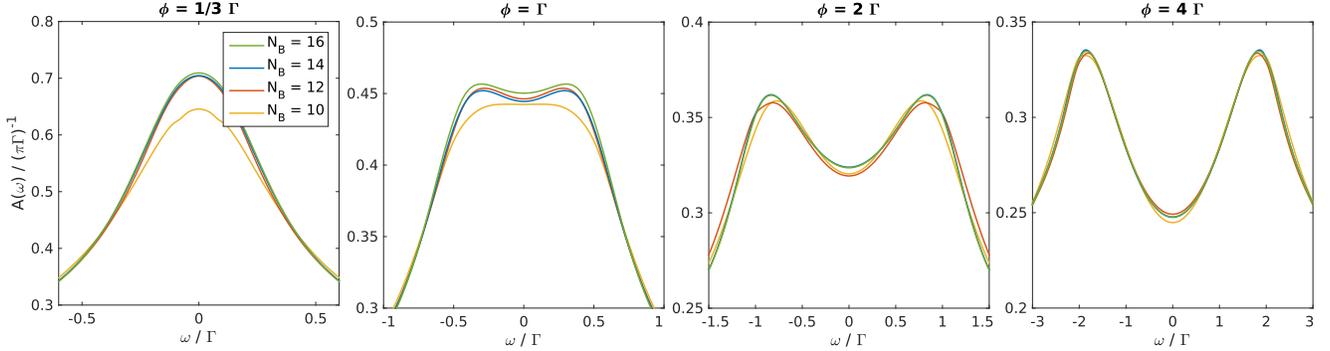}
\caption{(Color online) Convergence of the spectral function as depicted in \fig{fig:Aw_phi0} with increasing $N_B$, i.e. for the flat band model with $U = 6\,\Gamma$ and $T = 0.05\,\Gamma$.} 
\label{fig:Aw_NB}
\end{center}
\end{figure*}

\section{Multidimensional optimization}\label{app:opt}
In order to achieve $\Daux \approx \Dph$, we optimize the bath parameters $\vv{E}$, $\vv{\ga1}$ and $\vv{\ga2}$. For this a suitable parametrization is chosen, which yields a unique set of matrices $\vv{E}$, $\vv{\ga1}$ and $\vv{\ga2}$ for every parameter vector $\vv{x}$. The mean squared error is quantified by a cost function 
\begin{equation}
\mathcal{C}(\vv{x})^2 =  \sum\limits_{\alpha\in\{R,K\}}\int\limits_{-\omega_c}^{\omega_c} \iim\{ \Delta^\alpha_\mathrm{ph}(\omega) - \Delta^\alpha_\mathrm{aux}(\omega;\vv{x}) \}^2 W(\omega) d\omega \,, 
\label{eq:costfunc}
\end{equation}
with a certain cut-off $\omega_c$ and weighting $W(\omega)$, which we take to be constant in the present paper.

A variety of strategies exists to find the optimal parameter set $\vv{x}_\mathrm{opt}$ which minimizes a cost function as stated above. In previous work, \tcite{do.nu.14}, we employed a gradient-based method with a large number of random starting points. Such a deterministic minimization works well for rather small problems, but becomes inefficient in the higher-dimensional case $\dim(\vv{x}) \gtrsim20$. It is then of great advantage to employ methods which are able to overcome local minima. Appropriate Monte Carlo (MC) sampling based methods are for instance simulated annealing, multicanonical simulations or parallel tempering (PT).~\cite{ki.ge.83,berg00,le.ch.94,hu.ne.96,ea.de.05,ka.tr.06} Especially a feedback-optimized version of the latter has proven to be useful for our purposes. For details we refer to \tcites{hu.ne.96,ea.de.05} and in the following we outline only briefly the implementation as used in this work.

In PT, also called replica exchange, one regards $\mathcal{C}(\vv{x})$ as an artificial energy, defines a set of artificial inverse temperatures $\beta_m$ and samples for each temperature from the Boltzmann distribution $p^m(\vv{x}) = 1/Z_m\exp\left(-\mathcal{C}(\vv{x})\beta_m\right)$. A replica $\vv{x}^m_1$ is assigned to each $\beta_m$ and updated through a Markov chain with the Metropolis-Hastings algorithm.~\cite{berg05} These MC-sweeps generate a sequence of $\vv{x}^m_l$, $l=1,2,\dots$, which are distributed according to $p^m(\vv{x})$. In addition, a swapping of replicas $\vv{x}^m_l$ and $\vv{x}^{m+1}_l$ for neighboring inverse temperatures $\beta_m$ and $\beta_{m+1}$ is proposed after a certain number of sweeps. Again, a Metropolis probability is used for the swaps
\begin{equation}
 q_l^{m,m+1} = \min\left(1,\exp\left(\Delta\mathcal{C}^m_l \left(\beta_m-\beta_{m+1}\right) \right)\right)\,,
 \label{eq:p_swap}
\end{equation}
with $\Delta\mathcal{C}^m_l = \left(\mathcal{C}(\vv{x}^m_l)-\mathcal{C}(\vv{x}^{m+1}_{l})\right)$. The set of $\beta_m$ in PT has the purpose that the low temperatures enable an efficient sampling of regions where $\mathcal{C}(\vv{x})$ is small and the exchange with higher temperatures avoids trapping in local minima. To allow for an expedient exchange of replicas, the set of $\beta_m$ needs to be adjusted. For our purposes we chose a feedback strategy which shifts the values $\beta_m$ in order to achieve that the swapping probability \eq{eq:p_swap} becomes constant with respect to $m$. This strategy may not be the best possible choice in general, cf. \tcite{ka.tr.06}, but enables a fast feedback and quickly adjusts to large changes in the values $\mathcal{C}(\vv{x}^m_l)$. In addition, we modified $ q_l^{m,m+1} \to \max( q_l^{m,m+1},q_\mathrm{th} )$ with a certain threshold probability ($q_\mathrm{th}\approx0.1$), to avoid that during a PT-run a separation into several temperature sets occurs, which do not 
exchange replicas efficiently. This may violate balance conditions for thermodynamic observables but does not affect the applicability to minimization problems. For the other PT-parameters we proceeded in the following way: In a single sweep each coordinate of $\vv{x}^m_l$ was updated once and 10 sweeps were performed before attempting a swap. Around $20-30$ inverse temperatures $\beta_m$ were used. 

In general, one cannot expect to find the optimal solution in a nontrivial high-dimensional problem, but with the PT algorithm as outlined above we obtain an $\vv{x}_\mathrm{min}$ which minimizes the cost function locally and may furthermore fulfill $\mathcal{C}(\vv{x}_\mathrm{min}) \approx \mathcal{C}(\vv{x}_\mathrm{opt})$ to good approximation. For the largest systems considered in this work, $N_B \gtrsim 14$, a good starting point was found to be important. For the case of tridiagonal $\vv{E}$, $\vv{\ga1}$ and $\vv{\ga2}$, a convenient choice is to make use of $\vv{x}_\mathrm{min}$ from the next smaller system with $N_B-2$.

\section{Convergence as a function of $N_B$}\label{app:conv_NB}
\Fig{fig:Fit} depicts two typical results of the optimization described in \app\ref{app:opt}, for $N_B=12$ and $N_B=14$. It is apparent that rapid convergence is achieved when increasing $N_B$. For low temperatures $T$ we find that the biggest error in $\Daux$ occurs in the retarded component at the positions of the chemical potentials $\mu_{L/R} = \pm\frac{\phi}{2}$, see $N_B=12$. This is a consequence of optimizing $\DRaux$ and $\DKaux$ simultaneously. For higher temperatures, for instance $T = 0.5\,\Gamma$, this effect is much less pronounced.

A brief analysis of the convergence behavior of the mapping procedure with increasing $N_B$ is given in \fig{fig:conv_N}. We present values of the cost function $\mathcal{C}$, \eq{eq:costfunc}, for different temperatures and bias voltages. In general one finds an exponential convergence $\mathcal{C} \propto \exp(-rN_B)$ to good approximation and the higher the temperature, the higher the rate of convergence $r$. By averaging over results for different $\phi$ we estimate a scaling of $r\propto T^\frac{1}{4}$. One can deduce from the order of magnitude of $\mathcal{C}$, that the calculations presented for $A(\omega)$ at $T = 0.02\,\Gamma$ (\fig{fig:Aw_SigR_phi_Lorentzian}) are not converged to the same accuracy as the ones at $T = 0.05\,\Gamma$ (\fig{fig:Aw_SigR_phi_T0_05}) or $T = 0.5\,\Gamma$ (\fig{fig:Aw_SigR_phi_T0_5}), and larger systems with $N_B \gtrsim20$ would be needed. However, the accuracy is comparable to the $N_B=12$ results for $T = 0.05\,\Gamma$, which already yielded qualitative correct 
behavior and quite accurate spectral data, see also \fig{fig:Aw_phi0} and \fig{fig:Aw_NB}. The influence of $\phi$ is nonmonotonic and strongly dependent on the particular density of states in the leads. For the situations considered in this work we find the tendency that larger $\phi$ result in larger values of $\mathcal{C}$. For a more detailed analysis of the scaling with temperature and the mapping procedure in general we refer to \tcite{arri.15u}.

For the flat band case with $T = 0.05\,\Gamma$ we present a more thorough investigation by comparing the spectral function in the interacting case $U = 6\,\Gamma$ for different numbers of bath sites in \fig{fig:Aw_NB}. As can be anticipated from the cost function $\mathcal{C}$ in \fig{fig:conv_N}, the cases $\phi = 1/3\,\Gamma$ and $\phi = 2\,\Gamma$ are well-converged for $N_B = 16$, which manifests itself also in $A(\omega)$. The case $\phi = \Gamma$ exhibits larger values of $\mathcal{C}$ and one can note more significant changes in $A(\omega)$. Interestingly, rather high values of $\mathcal{C}$ are obtained for $\phi = 4\,\Gamma$, but nevertheless, the spectral function converges nicely. As discussed above for \fig{fig:Fit}, the largest errors in $\Daux$ correspond to short-scaled oscillations in $\DRaux$. These errors are likely to be averaged out once the spectral function exhibits rather broad features. This is exactly the case for higher bias voltages where the Kondo effect is strongly suppressed. 
On the whole, when inspecting \fig{fig:Aw_NB} and also \fig{fig:Aw_phi0}, one can note a slightly nonsmooth convergence with $N_B$, especially close to the Kondo regime for low $\phi$. This can be accounted to abrupt changes of spectral weight in $\DRaux$ around $\omega=0$, when changing $N_B$. One possibility to suppress this effect is to adjust the weighting function $W(\omega)$ in \eq{eq:costfunc} accordingly. However, this is most probably only of importance when aiming to achieve even higher accuracies in $A(\omega)$ and with the choice $W(\omega)=1$, one can regard the calculations presented in this work as unbiased and accurate over the whole $\omega$-domain.

\bibliography{mpsAMEA.bbl}{}

\begin{thebibliography}{138}
\bibitem[{foo({\natexlab{a}})}]{footnote_nrgTk}
\bibinfo{note}{We estimate $T_K$ based on a NRG calculation.\cite{footnote_nrg}
  A common definition related to the linear response conductance $G(T)$ is
  used: $G(T_K) = G(0)/2$.\cite{bu.co.08,os.zi.13,ha.we.14}}.
\bibitem[{foo({\natexlab{d}})}]{footnote_gamma}
\bibinfo{note}{As in previous work,\cite{do.nu.14} we set all couplings in
  $\ga1$ and $\ga2$ which connect to the impurity site to zero. This ensures
  that $\Daux$ approaches zero for $\omega \to \pm\infty$.}
\bibitem[{foo({\natexlab{e}})}]{footnote_Ivac}
\bibinfo{note}{The "tilde conjugation rules"\cite{dz.ko.11} are here
  $c_j\ket{I} = -i \tilde{c}_j\ket{I}$ and $c^\dagger_j\ket{I} = -i
  \tilde{c}^\dagger_j\ket{I}$.}
\bibitem[{foo({\natexlab{f}})}]{footnote_G_AB}
\bibinfo{note}{Steady state correlation functions for the case $t<0$ can be
  obtained via $\braket{A(t)B}_\infty^\dagger = \bra{I}B^\dagger
  e^{-Lt}A^\dagger \ket{\rho_\infty}$.}
\bibitem[{foo({\natexlab{g}})}]{footnote_mps}
\bibinfo{note}{The MPS matrices for the first and the last site are row and
  column vectors to be precise.}
\bibitem[{foo({\natexlab{h}})}]{footnote_splitting}
\bibinfo{note}{For the second-order splitting we used in particular the method
  of Strang (also known as Trotter-Suzuki) and of
  McLachlan,\cite{stra.68,mc.at.92,wa.ma.12} where the latter was
  advantageous.}
\bibitem[{foo({\natexlab{i}})}]{footnote_timeevo}
\bibinfo{note}{The particular initial state is not of importance, only the
  appropriate half-filling of the system in order to arrive at
  $\ket{\rho_\infty}$. A convenient choice is to use $\ket{I}$, for instance.}
\bibitem[{foo({\natexlab{j}})}]{footnote_G_phsymm}
\bibinfo{note}{Note that in the particle-hole symmetric case it is sufficient
  to calculate a single Green's function $G^<$ or $G^>$.}
\bibitem[{foo({\natexlab{k}})}]{footnote_EDcheck}
\bibinfo{note}{It was even found that the relative error saturates over time.
  This is likely due to the exponential damping of excited states
  $c^{(\dagger)}_{i\sigma}\ket{\rho_\infty}$ and the resulting contraction.}
\bibitem[{foo({\natexlab{l}})}]{footnote_Lorentzian_Tphi}
\bibinfo{note}{$T = 0.02\,\Gamma$ is small enough to observe two distinct
  "Fermi edges" in $\DKaux$ already for $\phi = 0.1\,\Gamma$ ($\lessapprox
  T_K$).}
\end{thebibliography}

\end{document}